# Revealing the origin of XMCD in an altermagnet via three-dimensional control of spins

Daire Mallon[1], Zixuan Wu[1], Jheng-Cyuan Lin[2], Ruiwen Xie[3], Bo Zhao[3], Charles Godfrey[1], Qing He[4], Lucia Iglesias[5], Pierluigi Gargiani[6], Manuel Valvidares[6], Peter Bencok[2], Francesco Maccherozzi[2], Larissa S I Veiga[2], Paul Steadman[2], Manuel Bibes[5], Hongbin Zhang[3], Paolo G Radaelli[1],[†‡] Hariom Jani[1].[†*]

[1]Clarendon Laboratory, Department of Physics, University of Oxford, Oxford, UK.
[2]Diamond Light Source, Harwell Science and Innovation Campus, Didcot, UK.
[3]Department of Materials and Earth Sciences, Technical University of Darmstadt, Darmstadt, Germany.
[4]Department of Physics, Durham University, Durham, UK.
[5]Laboratoire Albert Fert, CNRS, Thales, Université Paris-Saclay, Paris, France.
[6]ALBA Synchrotron Light Facility, CELLS, Cerdanyola del Vallés, Spain.

† These authors contributed equally in leading this work
‡ paolo.radaelli@physics.ox.ac.uk
* hariom.jani@physics.ac.uk

## Abstract

Altermagnets are an emerging class of collinear antiferromagnets that exhibit unconventional spin-polarised electronic bands, potentially unlocking new functionalities that do not rely on spin-orbit coupling (SOC). Experimental signatures traditionally associated with spin polarisation, like X-ray magnetic circular dichroism (XMCD), are thus being used as a validation of altermagnetism. However, unlike altermagnetic spin-splitting, these responses require SOC and are not invariant under spin-space rotations. This brings into question the extent to which they can be considered direct signatures of altermagnetism. Here, we exploit the g-wave altermagnet $\alpha$-$Fe_2O_3$ to demonstrate that XMCD is governed precisely by the spin-direction-induced symmetry breaking that altermagnetic spin groups are designed to ignore. Strikingly, the XMCD is highly anisotropic and is decoupled from the weak magnetic canting. We show that this anomalous XMCD can be described by on-site Faraday tensors capturing the locally uncompensated spin-orbital anisotropies – a scenario that can be applied to other altermagnets. Leveraging this, we reconstruct complete vectorial maps of nanoscale textures in $\alpha$-$Fe_2O_3$ thin films, including domain walls and topological solitons, which are promising for building future spintronics and magnonics devices.

# Main

Collinear antiferromagnets have recently emerged as platforms to harness unconventional electronic, spintronic and optical responses.[1-6] This is largely driven by the realisation that altermagnets – the class of collinear antiferromagnets that break $\mathcal{PT}$ and $\mathcal{T}\tau$ symmetries ($\mathcal{P}$: inversion, $\mathcal{T}$: time-reversal, $\tau$: translation) – harbour giant spin polarisation in the electronic bands that is primarily of non-relativistic exchange origin. This has motivated a symmetry-based reclassification of collinear magnets in terms of the magnetic spin groups (SGs), which describe symmetries that retain full Heisenberg SO(3) rotational invariance in spin space.[1,2] When applied to (quasi-)collinear magnets, this results in a tripartite distinction between ferro/ferrimagnets, altermagnets and non-altermagnetic antiferromagnets, based on the presence or absence of alternating spin-splitting in momentum space.

This framework can also be rigorously extended to macroscopic responses, based on whether or not they are related to spin-splitting by symmetry. In fact, one can cleanly distinguish between '*co-rotating*' and '*non-co-rotating*' tensorial properties, as defined rigorously in the Methods section. Here, co-rotating (covariant) tensors follow the SG classification and leave the magnitude of the corresponding physical properties invariant under spin-space rotations, as expected in the absence of SOC. Examples of such responses in altermagnets include the spin-splitter effect, piezomagnetism, and linear magnetic birefringence.[5-11] Alternatively, altermagnets can also host many non-co-rotating responses, activated by $\mathcal{PT}/\mathcal{T}\tau$ symmetry breaking, which instead follow the magnetic point group (MPG) classification, and generally require SOC to exist. Such non-co-rotating responses depend not only on the magnetic order pattern (bipartite in a collinear antiferromagnet), but also on how the Néel vector is oriented in the crystal. Most conventional properties of magnetism, including weak magnetisation, X-ray magnetic circular dichroism (XMCD), magneto-optical Kerr effect (MOKE), and anomalous Hall effect, do not admit a co-rotating component in altermagnets.[12-21]

This distinction raises a central question: what is the relationship between altermagnetism and non-co-rotating properties like XMCD? In ferromagnets and most ferrimagnets, XMCD is dominated by an isotropic contribution proportional to the magnetisation, which is co-rotating with the order parameter and is governed by the sum rules.[22-25] By contrast, in compensated magnets, spin contributions mostly cancel, and any XMCD response must arise from SOC and spin-orbital anisotropies not captured by a spherically symmetric description.[24,26-28] This leads to peculiar XMCD signatures, which recent studies have interpreted as straightforward proxies for altermagnetism.[12-14,17-19]

Here, by studying the XMCD response of the classic altermagnet α-$Fe_2O_3$, we demonstrate that this interpretation is incorrect. Compared to other recently studied systems (such as MnTe, $NiF_2$, and $BiFeO_3$),[12-14,18,19] α-$Fe_2O_3$ offers two distinctive advantages: three-dimensional control of the Néel vector with access to multiple MPGs within a single SG configuration, and the decoupling of the XMCD vector from the weak magnetisation. Leveraging this, we firmly establish the non-co-rotating character of XMCD, and show that the non-trivial anisotropy of XMCD does not map onto the altermagnetic SG classification. We also show that this 'anomalous' XMCD is not related to any symmetries that are peculiar to altermagnets. Rather, it is the result of the near-perfect sublattice compensation, which suppresses the isotropic spin contributions and unmasks the weak terms due to local anisotropy. Importantly, this local anisotropy is not pre-built in the paramagnetic crystal structure, but is activated by a local magneto-elastic-like response to the direction of the Néel vector. Finally, we exploit this anomalous XMCD together with X-ray magnetic linear dichroism (XMLD)[29,30] to reconstruct complete vector maps of the altermagnetic domain structure in α-$Fe_2O_3$.

### Magnetic symmetries of α-$Fe_2O_3$

α-$Fe_2O_3$ crystallises in a trigonal structure (space group $R\bar{3}c$) composed of face-sharing distorted $FeO_6$ octahedra (Figure 1). Its symmetry group comprises one 3-fold roto-inversion axis ($c$-axis), three 2-fold axes ($a$-axes), an inversion centre, and three vertical glide mirrors. Spins are ferromagnetically aligned within the basal planes and antiferromagnetically stacked along $c$-axis, with a $(+,-,-,+)$ configuration, such that parallel spins are connected by inversion.[31-33] This structure can be reduced to two effective, oppositely aligned sublattices associated with adjacent octahedra (Figure 1), and the Néel vector is defined as $\boldsymbol{L} = \boldsymbol{M}_1 - \boldsymbol{M}_2$, where the indices 1 and 2 refer to the two sublattices.

The magnetic structure of α-$Fe_2O_3$ breaks both $\mathcal{PT}$ and $\mathcal{T}\tau$ symmetries. It is classified as a bulk-type *g-wave* altermagnet with spin point group $\bar{3}^1m^2$ (spin space group $R\bar{3}^1c^2$) where 3-fold rotations and inversion map the spins onto the same magnetic sublattice, while 2-fold rotations and $c$-glides transpose the sublattices.[1,2,11,34] At the band structure level, and in the absence of SOC, this symmetry results in the formation of alternating spin-split electronic bands in momentum space, partitioned by 4 nodal surfaces reflecting the trigonal crystal symmetry.[15,17,34] The rank-5 tensor describing the altermagnetic spin splitting strictly co-rotates with $\boldsymbol{L}$, leading to the magnitude of the spin splitting being independent of the orientation of $\boldsymbol{L}$ relative to the lattice.[6]

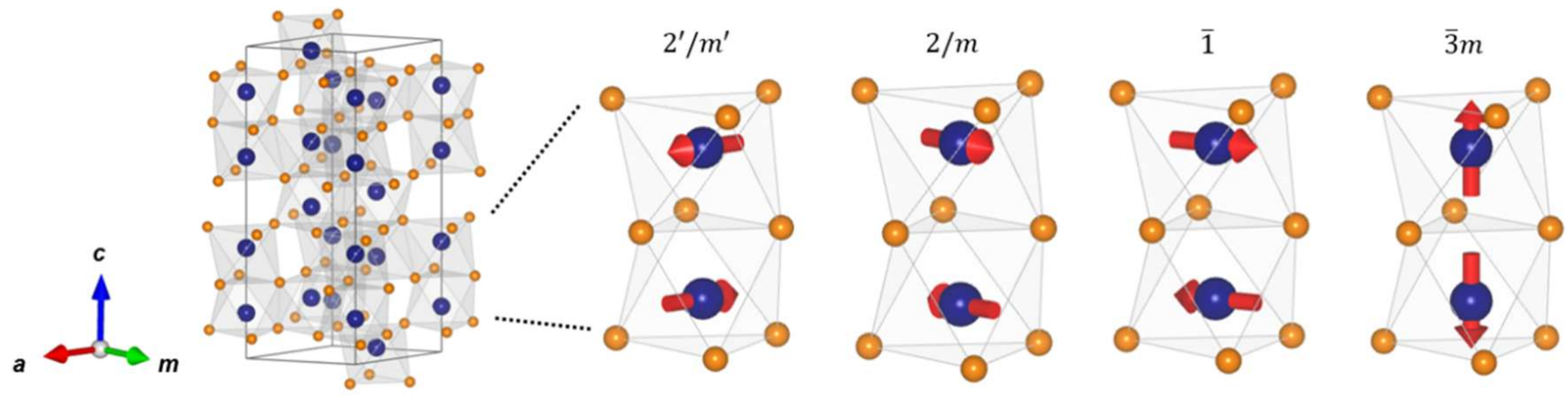

***Figure 1 | The corundum crystal structure of α-$Fe_2O_3$ and the accessible magnetic point groups.*** *The MPGs include $2'/m'$ (spins along the 2-fold axis), $2/m$ (spins orthogonal to the 2-fold axis), $\bar{1}$ (spins in a low symmetry direction) and $\bar{3}m$ (spins along the high-symmetry 3-fold axis). The Fe-spins are visualised within the adjacent face-sharing $FeO_6$ octahedra, showing the opposite directions of the magnetic sub-lattices $\boldsymbol{M}_1$ and $\boldsymbol{M}_2$. The a-axis and m-axis are orthogonal and span the basal plane.*

The inclusion of relativistic SOC fundamentally modifies this picture. The overall symmetry is lowered by coupling crystal-space and spin-space symmetries, leading to the well-known classification of the magnetic structures in terms of MPGs. α-$Fe_2O_3$ hosts four distinct MPGs, $\bar{3}m$, $2'/m'$, $2/m$ and $\bar{1}$, depending on the orientation of $\boldsymbol{L}$, see Figure 1. Crucially, α-$Fe_2O_3$ is only weakly anisotropic, and undergoes a spin reorientation 'Morin' transition at $T_M \sim$ 260 K where the axial anisotropy changes sign from easy-axis ($T < T_{\mathrm{M}}$) to easy-plane with weak trigonal anisotropy ($T > T_{\mathrm{M}}$).[29,35,36] Because this transition is highly sensitive to external perturbations (strain, field, temperature, doping),[29,31,35,37-41] $\boldsymbol{L}$ can be continuously reoriented in three dimensions, offering a unique opportunity to investigate the dependence of non-co-rotating responses on the direction of $\boldsymbol{L}$ across multiple MPGs.

### Non-co-rotating magnetic dipole-like vectors in α-$Fe_2O_3$

A key consequence of the SOC-induced symmetry reduction is the emergence of $\mathcal{T}$-odd axial vectors (magnetic dipole-like) corresponding to physical observables that are not co-rotating with $\boldsymbol{L}$, and whose existence and orientation are dictated by the MPG rather than the SG. By symmetry, such $\mathcal{T}$-odd axial vectors must be: (i) confined to the glide plane for $2'/m'$, (ii) orthogonal to the glide plane for $2/m$, (iii) unconstrained for $\bar{1}$, and (iv) forbidden for $\bar{3}m$.

Among such vectors, the best known by far is the weak canted magnetisation, $\boldsymbol{M} = \boldsymbol{M}_1 + \boldsymbol{M}_2$, which is quite small (<5 $\times 10^{-3}$ $\mu_B$/atom) but measurable even in thin films.[31-33,35,42] Because $\boldsymbol{M}$ couples the magnetic structure directly to external fields, it enables full three-dimensional control of antiferromagnetism, including in-plane rotation and spin-flop transitions. To an excellent approximation, the weak magnetisation is given by the expression $\boldsymbol{M} = \boldsymbol{L} \times \boldsymbol{D}$, with the Dzyaloshinskii-Moriya vector $\boldsymbol{D} \parallel \hat{\boldsymbol{c}}$, so that $\boldsymbol{M}$ is predominantly orthogonal to the $c$-axis, even when a $c$-axis component is allowed (e.g. MPG $2'/m'$).[31] As implied by this formula, $|\boldsymbol{M}|$ is largely independent on the in-plane orientation of $\boldsymbol{L}$. Other $\mathcal{T}$-odd vectors (referred to as $\boldsymbol{T}$ hereafter) are also allowed, such as those defining the antisymmetric part of the X-ray (XMCD) and visible (MOKE) optical conductivity. Importantly, for MPGs $2'/m'$ and $\bar{1}$, $\boldsymbol{T}$ and $\boldsymbol{M}$ need not be collinear.

Within this framework, XMCD in α-$Fe_2O_3$ is governed by an 'effective' magnetic dipole vector $\boldsymbol{T}$, which is related to the antisymmetric part of the energy-dependent optical conductivity averaged over both magnetic sublattices, $T_i(E) = \epsilon_{ijk}\, \beta_{jk}^{\mathrm{avg}}(E)$. Hence, $\boldsymbol{T}$ is directly connected to the XMCD intensity via the relation $I^{\mathrm{XMCD}} \propto \boldsymbol{T} \cdot \boldsymbol{k}$, where $\boldsymbol{k}$ is the X-ray wavevector. The XMCD $\boldsymbol{T}$ vector can be expanded in odd powers of $\boldsymbol{L}$ and $\boldsymbol{M}$ and naturally vanishes if $\boldsymbol{L}$ is along the $c$-axis. If $\boldsymbol{L}$ is in the basal plane, the lowest-order contributions (1st and 3rd orders) to $\boldsymbol{T}$ are given by (see SI):

$$\boldsymbol{T}^{(1)} = -f_{132}\boldsymbol{M} + f_{131}(\boldsymbol{L} \times \hat{\boldsymbol{c}}) \quad [1]$$

$$\boldsymbol{T}^{(3)} = -\frac{3}{2}c_{1312}|\boldsymbol{L}|^2\boldsymbol{M} + \frac{1}{2}c_{1311}|\boldsymbol{L}|^2(\boldsymbol{L} \times \hat{\boldsymbol{c}}) + \frac{1}{4}|\boldsymbol{L}|^2(c_{1211}|\boldsymbol{L}| - 3c_{1212}|\boldsymbol{M}|)\cos 3\varphi\, \hat{\boldsymbol{c}} \quad [2]$$

where $\varphi$ is the azimuthal angle between $\boldsymbol{L}$ and the $a$-axis, and $f_{ijk}$ and $c_{ijkl}$ are components of the linear and cubic Faraday tensors (for simplicity, we have dropped the quadratic and cubic contributions in $\boldsymbol{M}$). Equations 1,2 are completely general for the corundum structure of α-$Fe_2O_3$ (hence, also valid for the MOKE vector), and were obtained by averaging the Faraday responses of two magnetic ions with crystallographic point-group symmetry 3 on (nearly) compensated magnetic sublattices. It is important to emphasise that, in spherical and cubic symmetries, all the co-efficients of Equations 1, 2 would be zero except for $f_{132}$ and $c_{1312}$ – in other words, $\boldsymbol{T}$ *would be strictly parallel to* $\boldsymbol{M}$ and would vanish for a perfectly compensated system.

Although these XMCD contributions in Equations 1,2 can originate from both $\boldsymbol{L}$ and $\boldsymbol{M}$, most of the terms are parallel to $\boldsymbol{M}$, since $\boldsymbol{M} \propto \boldsymbol{L} \times \hat{\boldsymbol{c}}$ (see above), with the only exception being the last term in $\boldsymbol{T}^{(3)}$ which defines a contribution *parallel* to the $c$-axis and varying as $\cos 3\varphi$. The $c$-axis component of $\boldsymbol{T}$ reaches an extremal value for MPG $2'/m'$ ($\varphi = n\pi/3$), reduces for MPG $\bar{1}$, and vanishes for MPG $2/m$ ($\varphi = \pi/2 + n\pi/3$). This, together with the observation that $\boldsymbol{T} = \boldsymbol{0}$ if $\boldsymbol{L} \parallel \hat{\boldsymbol{c}}$, are in complete agreement with the MPG analysis, and confirm that all components of the $\boldsymbol{T}$ vector depend on the orientation of $\boldsymbol{L}$ and, hence, are non-co-rotating.

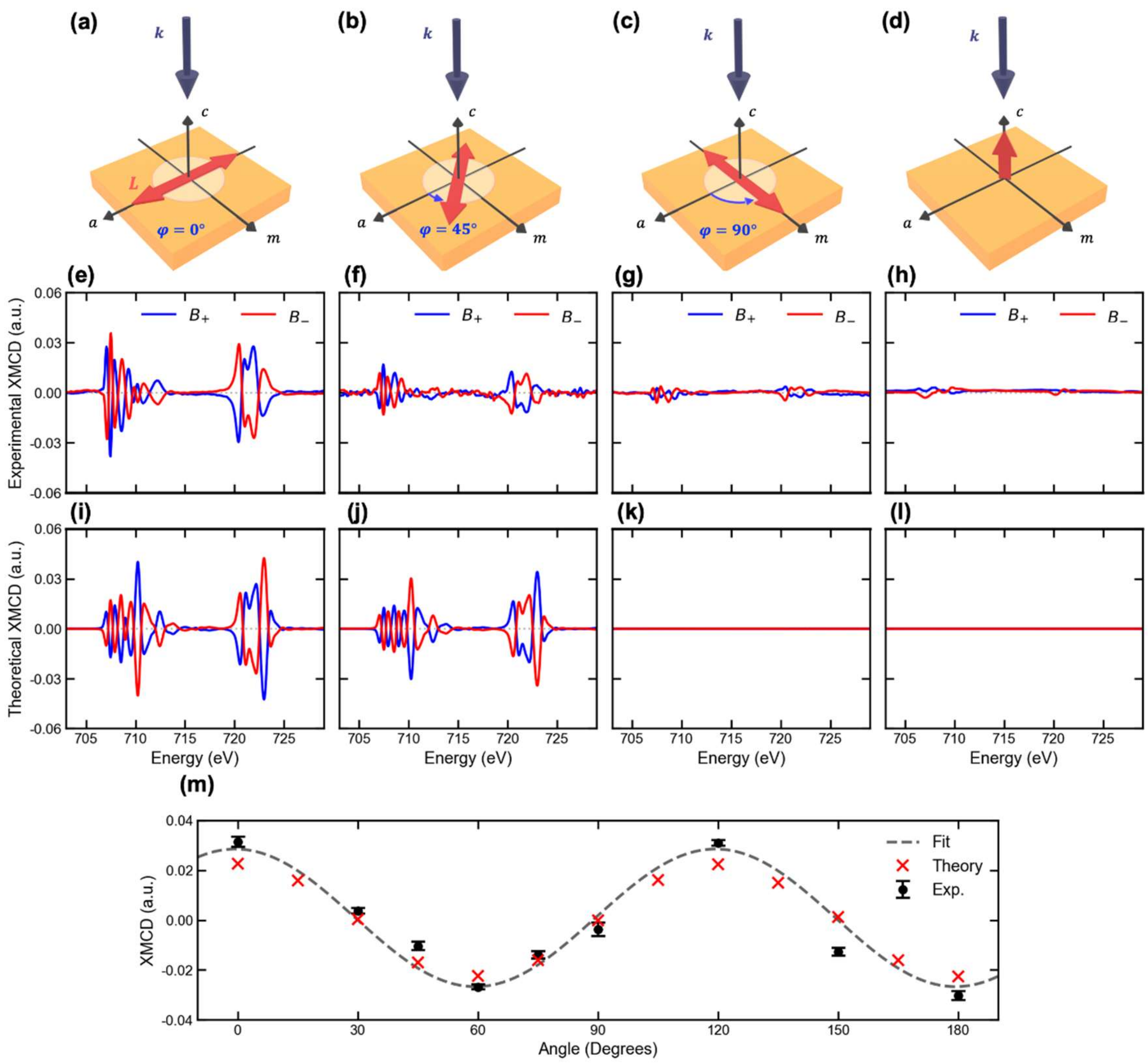

***Figure 2 | Fe L-edge XMCD spectra with X-ray k-vector along the c-axis. (a-d)*** *Experimental geometries (normal incidence) and* ***L*** *orientations (defined by antiparallel sublattices - blue arrows) of MPGs present within α-Fe₂O₃, (a)* $2'/m'$*, (b)* $\bar{1}$*, (c)* $2/m$ *and (d)* $\bar{3}m$ *with respect to crystal axes.* ***(e-h)*** *Measured and* ***(i-l)*** *calculated XMCD spectra for four accessible MPGs. Spectra in (e-g) were measured in* the *total electron yield* mode *at T = 300 K. An orthogonal magnetic field was applied to align* ***L*** *parallel/antiparallel to the crystal directions indicated (blue/red). The out-of-plane* ***L*** *orientation in (h) was achieved by cooling below* $T_M$ *(T = 12 K).* ***(m)*** *XMCD intensity at* $L_2$*-edge peak (722.25 eV) as a function of the in-plane angle φ of* ***L*** *from the a-axis. Measurements were performed in both total electron yield (TEY) and fluorescence yield (FY)* modes *at T = 300 K. Error bars represent the standard error of the mean, calculated from the distribution of XMCD values obtained from individual scans (left-right polarisation) at the fixed energy.*

## Vectorial XMCD: Out-of-plane and in-plane components

We first measured the out-of-plane component of $\boldsymbol{T}$, $T_z$, using normal-incidence XMCD on a c-cut α-$Fe_2O_3$ crystal (Figure 2). We oriented $\boldsymbol{L}$ via the weak canted moment using an in-plane magnetic field. Above $T_{\mathrm{M}}$, we obtained a strong XMCD signal when $\boldsymbol{L}$ lies along the 2-fold $a$-axis (Figure 2e). The spectra exhibit characteristic oscillations at the $L_3$ and $L_2$ edges, which integrate to ~ 0 at each edge and reverse sign upon reversal of the magnetic field (and hence of $\boldsymbol{L}$). This configuration corresponds to the MPG $2'/m'$, for which $T_z$ is allowed by symmetry, and its magnitude is maximum as per Equation 2. When $\boldsymbol{L}$ is aligned along the $m$-axis, the XMCD signal nearly vanishes (Figure 2g), consistent with MPG symmetry $2/m$, and also with Equation 2, which forbids an out-of-plane component. Residual signals are attributed to small misalignments of the external field and $\boldsymbol{L}$ with the crystal axes. For intermediate orientations of $\boldsymbol{L}$, corresponding to MPG $\bar{1}$, XMCD reappears with reduced magnitude and reverses sign with the field (Figure 2f), consistent with the symmetry constraints on $\boldsymbol{T}$. Finally, below $T_{\mathrm{M}}$, when $\boldsymbol{L}$ is aligned along the $c$-axis (MPG $\bar{3}m$), XMCD vanishes entirely (Figure 2h), consistent with the fact that here any $\mathcal{T}$-odd axial vector is forbidden.

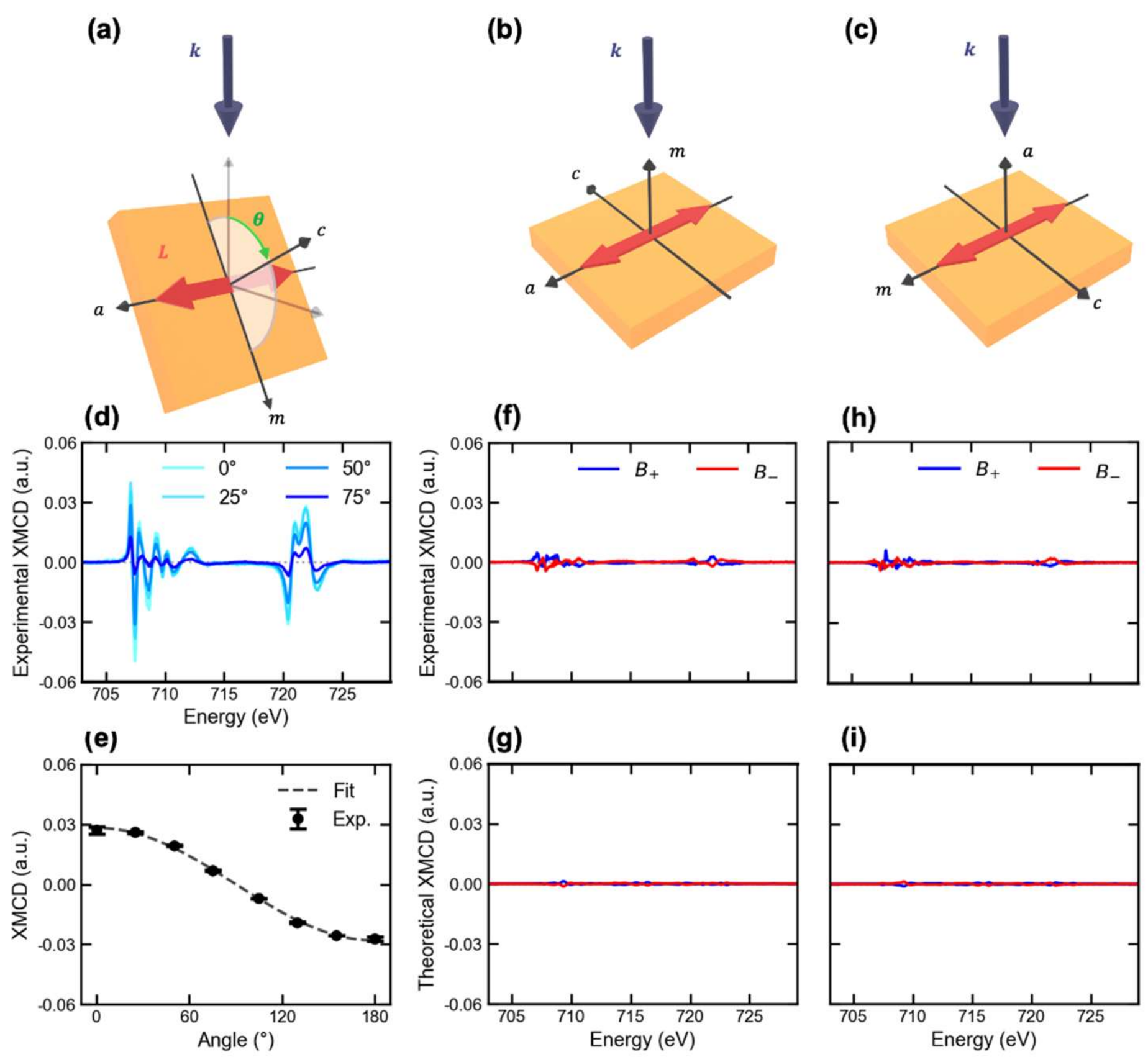

*Figure 3 | Fe L-edge XMCD spectra with X-ray k-vector oblique and orthogonal to the c-axis.* ***(a-c)*** *Experimental geometries (oblique and normal incidence) and* $\boldsymbol{L}$ *orientations (defined by antiparallel sublattices - blue arrows) used to measure: the (a) evolution of the* $T_z$ *component with rotation of the sample normal, and the (b,c)* $T_{y,x}$ *components of the XMCD vector.* ***(d)*** *Modulation of the XMCD signal under increasing angle of incidence,* $\theta$*, from the normal, at T = 300 K.* ***(e)*** *XMCD intensity at* $L_2$*-edge peak (722.25 eV) as a function of* $\theta$*. Error bars represent the standard error of the mean, calculated from the distribution of XMCD values obtained from individual scans (left-right polarisation) at the fixed energy.* ***(f,h)*** *Measured and* ***(g,i)*** *calculated normal incidence XMCD signals for (f,g) m-cut and (h,i) a-cut crystals. An orthogonal magnetic field was applied to align* $\boldsymbol{L}$ *parallel/antiparallel to the crystal directions indicated (blue/red).*

The evolution of the $T_z$ component of the XMCD signal and can be obtained through the rotation of $\boldsymbol{L}$ in the basal plane, tracked at fixed energies (see Figure 2m and Extended Data Figure 1). The signal exhibits a clear 120° periodicity ($\propto\ cos\ 3\ \varphi$) in complete agreement with Equation 2. The observed modulations directly confirm that this component of XMCD is not a co-rotating property defined by the SG framework. Rather, the XMCD vector is in a fixed direction orthogonal to $\boldsymbol{L}$ , and its magnitude depends on the orientation of $\boldsymbol{L}$ in the crystal.

Our experimental spectra are reproduced by calculations based on density functional theory and multiplet ligand field theory (see Figure 2 and Methods).[15] The calculations capture both the energy-dependent spectral structure and the 3-fold azimuthal dependence, confirming that the observed XMCD originates from the underlying symmetry constraints rather than extrinsic effects or defects. Minor discrepancies in magnitude are attributed to the values of model parameters, e.g., crystal fields are usually underestimated in DFT calculations.[43]

Next, we investigated the in-plane projected components $T_{x,y}$, orthogonal to the $c$-axis. We emphasise that this component includes the 'conventional' XMCD, which is allowed in the spherical approximation. Oblique-incidence XMCD measurements on $c$-cut crystals are obtained by rotating the polar angle between the $c$-axis and $\boldsymbol{k}$, see Figure 3a. Here, the field is applied to align $\boldsymbol{L}$ along the $a$-axis, which serves as the axis of rotation. This geometry corresponds to probing the vector component $T_z\ \cos\theta + T_{x,y}\sin\theta$, where $\theta$ is the rotation angle. Upon rotation from normal to grazing incidence, we find that the spectral signature reduces, with no sign reversal or significant change in its structure and no new component appearing. This is consistent with a gradually decreasing contribution of the $T_z$ component, following a $\cos\theta$ angular dependence, while $T_{x,y} \approx 0$. We obtained similar results from grazing incidence experiments on the other samples.

To confirm this surprising result, and suppress the contribution from $T_z$ entirely, we then measured normal-incidence XMCD on $a$-cut and $m$-cut crystals (Figure 3b,c). Here, the field is applied out-of-plane to ensure that $\boldsymbol{L}$ is along the $a$-axis (MPG $2'/m'$) and $m$-axis (MPG $2/m$), respectively. In both cases, XMCD is negligible, which is consistent with theoretical calculations and confirms that $T_{x,y} \approx 0$. Any residual signal can again be attributed to either noise or a residual $T_z$ component due to sample misalignment during mounting.

### XMCD of the canted magnetism

To activate the 'conventional' XMCD response associated with spin canting, we perform grazing-incidence XMCD measurements on $c$-cut α-$Fe_2O_3$ under large in-plane magnetic fields, which enhance the sublattice canting $\boldsymbol{M}$, see Figure 4. In addition to the dominant XMCD signal discussed above, we observe a smaller field-dependent contribution that is negligible at small fields but becomes pronounced at high fields (Extended Data Figures 2). This is extracted by subtracting a low-field reference spectrum from the high-field XMCD signal, thereby isolating the field-induced response. This difference signal (ΔXMCD) was measured for all three relevant MPGs, and in all cases, its magnitude increases linearly with the applied field, consistent with the progressive canting of the sublattices and the corresponding increase of $\boldsymbol{M}$, see Figure 4e.

Crucially, this '$\boldsymbol{M}$-related' XMCD contribution is isotropic, depending only on the projection of $\boldsymbol{M}$ onto $\boldsymbol{k}$ and shows no dependence on the crystallographic orientation. This XMCD signature can hence be ascribed to the 1st-order tensor, scaling linearly with $\boldsymbol{M}$ (Equation 1). As previously discussed, this is characteristic of a conventional XMCD response, arising from a spinful monopolar (spherical) contribution analogous to that observed in ferro-/ferrimagnets. It therefore provides a direct contrast to the 'anomalous' $c$-axis XMCD, which is an anisotropic, higher-order process, constrained by MPGs.

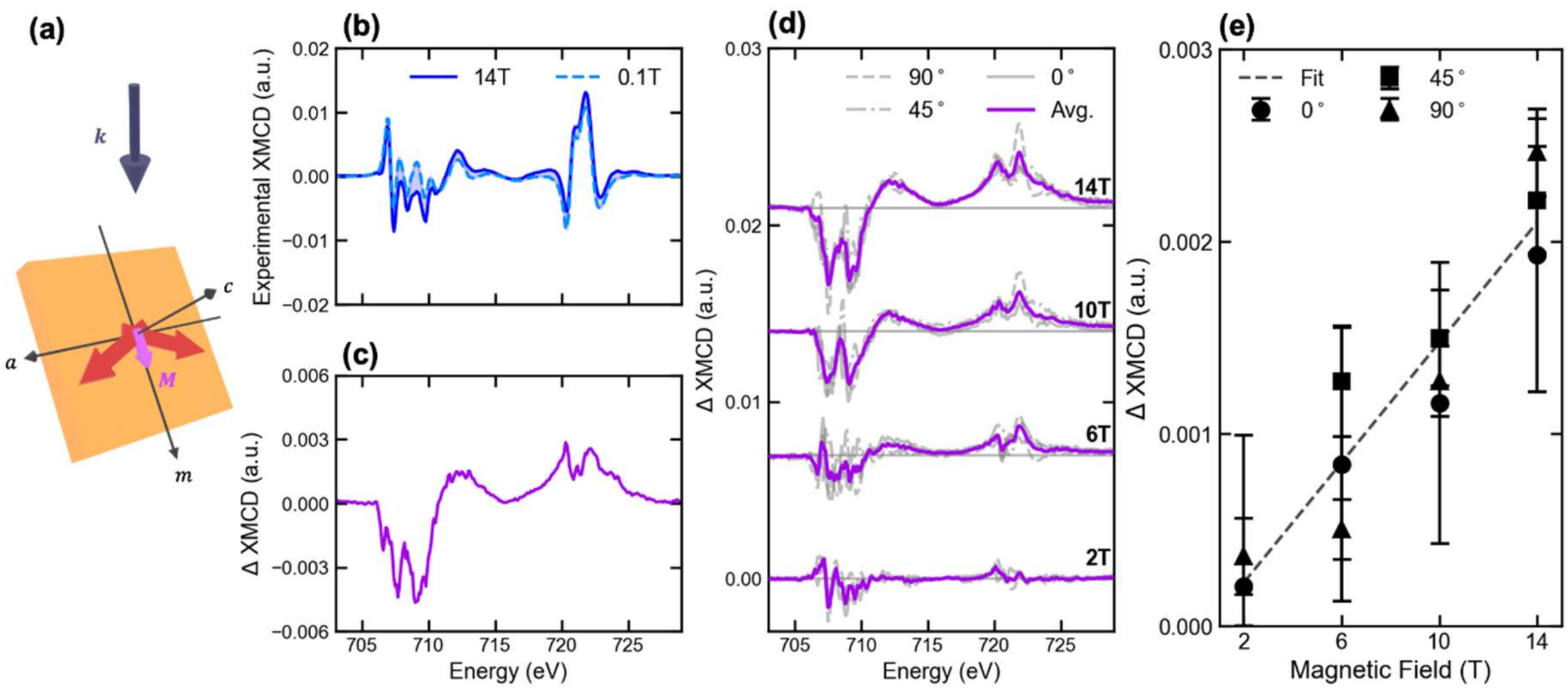

*Figure 4 | Fe L-edge XMCD signature of the weak canted moment.* ***(a)*** *Experimental grazing-incidence (75°) geometry used to isolate the XMCD response of the weak moment (purple arrow), emerging from the canting of antiparallel sublattices (blue arrows).* ***(b,c)*** *XMCD spectra measured for c-cut crystal above $T_M$ (T = 300 K), under (b) positive (blue) and (c) negative (red) applied magnetic fields parallel to the X-ray beam. Dashed and solid lines correspond to field strengths of 0.1 T and 14 T, respectively.* ***(d)*** *Difference XMCD spectra under increasing applied magnetic fields for three in-plane* ***L*** *orientations* corresponding to the MPGs $2'/m'$, $2/m$, $\bar{1}$. ***(e)*** *XMCD intensity at the $L_2$-edge (720 eV) of the* canted moment *as a function of applied magnetic field. Error bars represent the standard error of the mean, calculated from the distribution of window-averaged XMCD values obtained from individual scans (left-right polarisation) at the fixed energy.*

### Connection to the general expression of *T*

We can compare these results to the tensorial expansion in Equations 1,2, and conclude that within experimental sensitivity, the anisotropic XMCD is dominated by the $c$-axis response ($f_{131}$, $c_{1311} \approx 0$, $c_{1211} \neq 0$). Moreover, the isotropic XMCD signal, scaling with $\boldsymbol{M}$, is only activated under large fields ($f_{132}$ and $c_{1312} \neq 0$). Overall, the XMCD response is dominated by a $\mathcal{T}$-odd axial vector $\boldsymbol{T}$ that is *parallel* to the $c$-axis and *perpendicular* to $\boldsymbol{L}$ and $\boldsymbol{M}$. The in-plane components of $\boldsymbol{T}$ are negligible even when allowed by symmetry.

This result establishes not only that XMCD is a non-co-rotating property, but also that its relationship with the altermagnetic structure of α-$Fe_2O_3$ is far more complex than it could be anticipated. For perfect compensation ($\boldsymbol{M} = 0$) XMCD would vanish at all energies in the spherical or cubic approximations. Hence, the 3-fold periodic signals we observe are indicative of local anisotropy and point to the essential role of SOC – precisely the kind of effects that the SG analysis is designed to ignore. Crucially, the absence of $\boldsymbol{L} \times \hat{\boldsymbol{c}}$ term and the cubic dependence on $\boldsymbol{L}$ indicate an even more fundamental role of SOC in the emergence of XMCD, as elaborated in the next section. We can also rule out any direct role of the momentum space spin-splitting near the Fermi energy, with and without SOC, in the XMCD signal. As shown in Extended Data Figure 3, the net spin polarisation is zero at all energies in the absence of SOC, while in the presence of SOC it is parallel to $\boldsymbol{M}$ and orthogonal to the $c$-axis.

## Microscopic origin of XMCD in α-$Fe_2O_3$

Microscopically, the $Fe^{3+}$ L-edge X-ray absorption arises due to electric dipole transitions from the half-filled ground state, $|2p^6\ 3d^5\rangle$, to the final state, $|2p^5\ 3d^6\rangle$, involving a core hole in the spin-orbit-split $2p$-levels and an extra $d$-shell electron. XMCD is sensitive to three magnetic contributions – the spinless orbital component, the isotropic spinful monopolar component, and the anisotropic spinful quadrupolar component.[22,24-27] These terms are directly connected to the Thole and Carra-Thole sum rules used for extracting the orbital ($m_L$) and effective spin angular momenta ($\mathrm{m_{eff}} = m_S + 7m_T$), where $m_L$, $m_S$ and $m_T$ are the ground-state expectation values of the orbital, spin-monopolar and spin-quadrupolar operators, respectively.

For a collinear antiferromagnet in the $3d^5$ configuration, the net orbital angular momentum and the quadrupolar components are completely quenched, while the spin-monopolar components are fully compensated between the sublattices if $\boldsymbol{M} = 0$. Hence, from the sum rules, one would expect an almost exact cancellation of the XMCD intensity integrated through each of the $L_3$ and $L_2$ edges.[24,26] However, the intensity at a given energy can be sizable, because the expectation values of the orbital and spin quadrupolar operators do not vanish for the photo-excited state, and the uncompensated contributions are almost exclusively along the $c$-axis.[26] The combination of all these effects explains the characteristic wiggling structure of the XMCD as a function of energy.

Since XMCD is a largely *local* probe, one can relate the coefficients in Equations 1,2 to the local structural distortions of the $FeO_6$ octahedra. Terms contributing to $T_{x,y}$, $f_{131}$ and $c_{1311}$, are permitted in the crystallographic point-group symmetries 3 and $\bar{3}$, but are both zero in the higher-symmetry groups 32, $3m$ and $\bar{3}m$. Hence, they must be related to an *axial* distortion (permitted in 3 and $\bar{3}$). However, such a distortion is not allowed structurally for an isolated $FeO_6$ octahedron, consistent with the fact that $T_{x,y}$ is not observed experimentally. By contrast, the dominant $T_z$-related term $c_{1211}$ is permitted in any local symmetry. This term is proportional to $\boldsymbol{L}^3$, and can be ascribed to a magneto-elastic-like response of the local site to the presence of a symmetry-breaking in-plane local moment. In other words, the deviation from spherical symmetry that dominates XMCD is not pre-baked in the crystal structure, but is induced by the interaction of the Néel vector with the local environment.

Our Faraday tensor framework can be extended to explain XMCD in other altermagnets. In particular, XMCD in MnTe has an identical dependence on the directions of $\boldsymbol{k}$ and $\boldsymbol{L}$, and, most likely, has the same origin as in α-$Fe_2O_3$.

## Complete vectorial imaging of nanoscale altermagnetic textures

We have established that XMCD in α-$Fe_2O_3$ is zero for $\boldsymbol{L}$ along the $c$-axis and has the form, $I^{\mathrm{XMCD}} \propto \boldsymbol{k} \cdot \boldsymbol{T} \propto \cos 3\varphi$ for $\boldsymbol{L}$ in the basal-plane. This expression becomes even more significant in the above-Morin phase, where weak in-plane anisotropy favours domains with MPG $2'/m'$, and hence extremal values of $I^{\mathrm{XMCD}}$ ($\varphi = n\pi/3$). For a given direction of $\boldsymbol{k}$, where $k_z \neq 0$, three of these domains ($\varphi = 0, 2\pi/3, 4\pi/3$) have equal $I^{\mathrm{XMCD}}$, while the $\mathcal{T}$-reversed counterparts ($\varphi = \pi, \pi/3, 5\pi/3$), have opposite $I^{\mathrm{XMCD}}$. Put simply, $I^{\mathrm{XMCD}}$ can distinguish between $\mathcal{T}$-reversed domains. It is also noteworthy that the XMCD signal is strong even in the unfavourable grazing geometry, provided $k_z \neq 0$. We have exploited this to perform full vectorial mapping of nanoscale altermagnetic textures using grazing-incidence X-ray photoemission electron microscopy (X-PEEM), see Figures 5,6 and Methods. Experiments were carried out on epitaxial α-$Fe_2O_3$ thin films grown on $c$-oriented α-$Al_2O_3$ substrates, which host a rich landscape of nanoscale spin textures[29,35,38,42] while exhibiting dichroic responses consistent with bulk crystals (see Extended Data Figure 4).

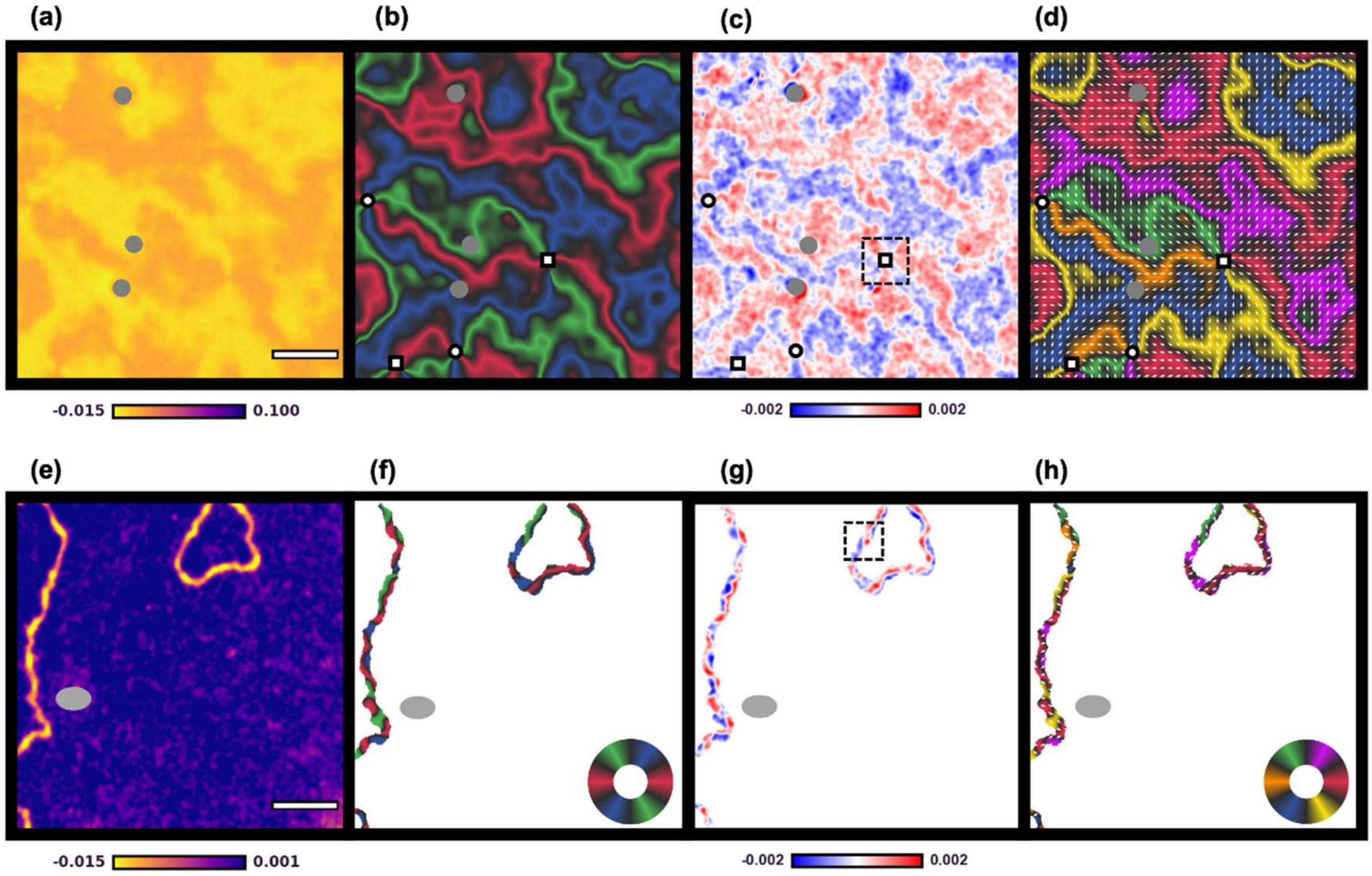

***Figure 5 | Fe L-edge X-ray vectorial imaging of altermagnetic domains in thin films of α-Fe₂O₃. (a-d)*** *X-PEEM imaging of the easy-plane phase (T = 400 K) and* ***(e-h)*** *easy-axis phase (T = 358 K). This includes: (a,e) XMLD X-PEEM images, (b,f) XMLD Néel vector maps, (c,g) XMCD-PEEM images and (d,h) the full vectorial maps, constructed as described in Methods. In (a,e), yellow and orange colours indicate orthogonal in-plane* $\boldsymbol{L}$ *orientations, while purple defines out-of-plane* $\boldsymbol{L}$*. In (b,f), R-G-B and black define in-plane* $\boldsymbol{L}$ *orientations along the a-axes and m-axes, respectively. The angular dependence is defined by the inset colour wheel. In (c,g), blue-white-red contrast corresponds to negative, zero and positive* $T_z$ *orientations. In (d,h), the colour wheel maps the complete 360° in-plane* $\boldsymbol{L}$ *orientations, removing the sign degeneracy of the Néel vector maps. The average* $\boldsymbol{L}$ *orientation of an area (~0.2 μm$^2$) is defined by a white arrow, with its direction corresponding to the inset colour wheel. Topological altermagnetic solitons, merons and antimerons, are highlighted as squares and circles, respectively. Spatial scale bar is 2 μm.*

We first imaged the magnetic state using XMLD, a $\mathcal{T}$-even probe that scales quadratically with $\boldsymbol{L}$. Utilising an established Néel vector mapping procedure that we developed previously,[29,30,35,44] we combine angle-dependent XMLD images to determine the orientation axes of $\boldsymbol{L}$, up to an arbitrary sign, see Methods and SI. We then measured XMCD to determine the local MPG and the relative orientation of $\boldsymbol{T}$. By combining XMLD and $\boldsymbol{T}$-XMCD, we reconstruct the full vectorial configuration of the altermagnetic order, including the sign of $\boldsymbol{L}$.[12]

Above $T_{\mathrm{M}}$, the sample is in an in-plane state, where XMLD phase reconstruction reveals a mosaic of trigonal domains (red-green-blue in Figure 5a,b) and their $\mathcal{T}$-reversed counterparts.[29,30,35,38,39] Domains are separated by 60° domain walls, where $\boldsymbol{L}$ remains confined in the basal plane. XMCD imaging reveals a strong contrast (red-blue in Figure 5c) that largely follows the in-plane domain structure. This contrast correlates with the local MPG: domains are defined by $2'/m'$ and $\bar{1}$, whereas domain walls corresponding to $2/m$. This real-space probe further confirms that XMCD is controlled by the local MPG. Reconstructing the full vector map of $\boldsymbol{L}$ (Figure 5d), we can resolve both its orientation and sign. The map confirms that neighbouring domains smoothly vary by a phase of not more than 60°, consistent with trigonal symmetry.

An illustrative state that captures all permissible MPGs simultaneously is a topological soliton. This is a nanoscale texture, hosted in the easy-plane state, where six domains intersect at a pinch-point.[29,35,38,45] Such textures consist of an out-of-plane core surrounded by an in-plane winding $\boldsymbol{L}$ that circulates along or opposite to the azimuthal angle in the basal plane (Figure 5b-d). These whirling solitons are characterised by a winding number, $w = \pm 1$, corresponding to altermagnetic merons and antimerons, which are topologically equivalent to half-skyrmions. Crucially, these solitons contain the MPG $\bar{3}m$ in the core, and $2'/m'$, $2/m$ and $\bar{1}$ in the winding region outside the core. As a result, the XMCD signal exhibits a characteristic 3-fold 'clover-leaf' pattern of $\boldsymbol{T}$, reflecting the spatial variation of $\boldsymbol{L}$ and the MPGs, as illustrated in Figure 6a,b. While some of these solitons can be isolated, they often appear in pairs, called bimerons (see Figure 5b-d).

Cooling the sample through $T_{\mathrm{M}}$, the in-plane domains gradually shrink and out-of-plane domains nucleate and expand to form the entire matrix, see Extended Data Figure 5. Below $T_{\mathrm{M}}$, the sample contains several $\mathcal{T}$-reversed out-of-plane domains (purple contrast in Figure 5e), partitioned by 180° anti-phase domain walls. XMLD vector reconstructions confirm that a large fraction of the image remains invariant under rotation (white contrast in Figure 5f), whereas the interior of the 180° walls contain an in-plane $\boldsymbol{L}$ that rotates freely within the basal plane (red-green-blue). In this phase, XMCD is absent within the bulk domains, consistent with the MPG $\bar{3}m$. A finite XMCD signal appears only within the domain walls. Mapping the nanoscale distribution of the full altermagnetic order (Figure 5h), we see that unlike the easy-plane state, we cannot distinguish $\mathcal{T}$-reversed out-of-plane domains. The variation of the XMCD emerges from the variation of the in-plane orientation of $\boldsymbol{L}$ along the length of the wall, as illustrated in Figure 6c,d.[29]

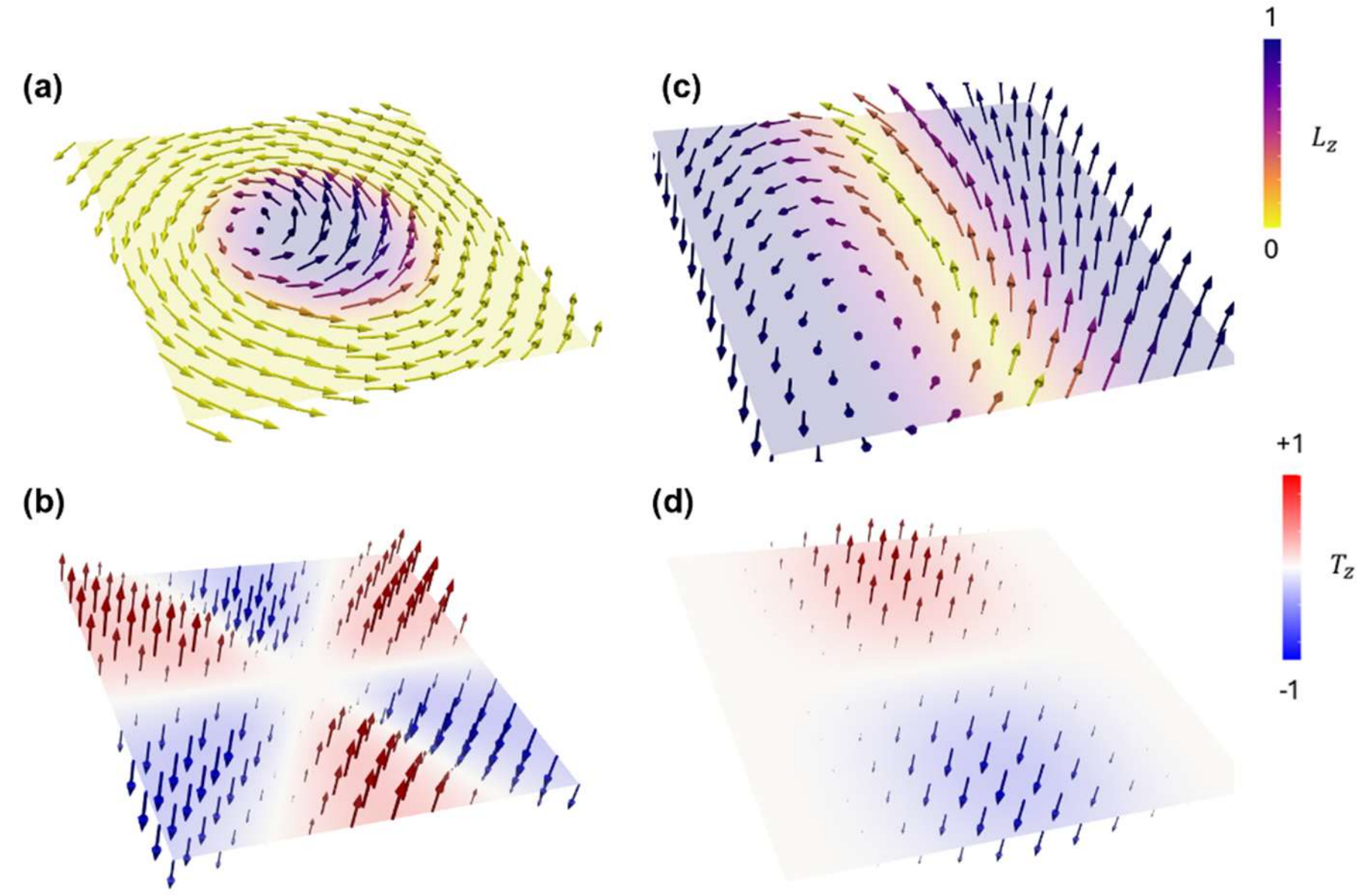

***Figure 6 | Illustration of the three-dimensional distribution of* L *and* T *in nanoscale altermagnetic textures.*** *Orientation of the local Néel vector* ***(a,c)****, and the corresponding XMCD vector* ***(b,d)*** *for a nanoscale meron in the easy-plane phase, and an anti-phase domain wall in the easy-axis phase. The XMCD vectors are shown for the dominant contribution, $T_z$, originating from the 3rd order Faraday tensor components described in Equation 2 and SI. Examples of both can be seen highlighted with dashed squares in Figures 5c and 5g, respectively.*

## Conclusions and Outlook

The defining characteristics of altermagnets, namely symmetry compensation, collinearity and the absence of $\mathcal{PT}$ and $\mathcal{T}\tau$ symmetries, enable a range of conventional and unconventional magnetic responses. These can be broadly divided into co-rotating and non-co-rotating properties. Co-rotating responses are symmetry-cognate with electronic spin-splitting and are unique to altermagnets. However, a wide variety of non-co-rotating responses, whilst also activated by $\mathcal{PT}/\mathcal{T}\tau$ symmetry breaking, require SOC to exist and are not linked to the altermagnetic symmetry framework.

Our results demonstrate that caution must be exercised in claiming non-co-rotating properties, like XMCD, as 'signatures of altermagnetism'. The principal reason for introducing the SG classification is the existence of co-rotating properties that are largely insensitive to the directional nature of $\boldsymbol{L}$. By contrast, we demonstrated that XMCD exists precisely in the opposite limit. In $\alpha$-$Fe_2O_3$, 'conventional' XMCD is quenched by the spin compensation, allowing weaker contributions from local site anisotropy to dominate. This anisotropy arises from SOC-driven coupling between $\boldsymbol{L}$ and the local environment, breaking the trigonal symmetry and activating Faraday tensor elements that remain uncompensated. As a result, the dominant XMCD vector lies along the high-symmetry axis and is decoupled from the weak magnetic canting. It is also unrelated to the altermagnetic spin-polarisation near the Fermi energy.

Building on this understanding, we showed that the XMCD can be used to distinguish time-reversed domains to create full vectorial images of nanoscale altermagnetic textures, including anti-phase domain walls and topological solitons. This will be useful for mapping the spatial distribution and ultrafast evolution of nanoscale altermagnetic states to design next-generation spintronic and magnonic devices.[1,2,45-47]

More broadly, our framework to distinguish co-rotating and non-co-rotating responses can be readily applied to classify a broad range of electronic, spintronic and optical phenomena across magnetic materials. Moreover, our Faraday tensor analysis can be used to explain the emergence of XMCD in other altermagnets and compensated antiferromagnets due to spontaneous or field-induced local anisotropies.

## Methods

### Theory

**Definition of co-rotating and non-co-rotating tensors:** In the absence of spin-orbit coupling, the internal SO(3) (or SU(2)) symmetry of the Hamiltonian translates into the fact that tensors describing allowed properties or responses, which can be called $T$, must be equivariant, or co-rotating, with the magnetic order parameter pseudovector $\boldsymbol{\Lambda}$. The precise statement here is: $T(R\boldsymbol{\Lambda}) = D(R)\, T(\boldsymbol{\Lambda})$, where $R \in SO(3)$ and $D(R)$ is a representation of $R$ appropriate for the rank of the response.

In a system with *collinear* magnetic ordering defined by the order parameter $\boldsymbol{\Lambda}$ (either the magnetisation or the Néel vector), one can construct co-rotating tensors of the separable form $T = \boldsymbol{\Lambda} \otimes B$ (or, in components $T_{k\,ij\ldots} = \Lambda_k B_{ij\ldots}$). Here, $B$ is a time-reversal-even tensor that is wholly invariant by spin-only rotations, and is defined purely by the *bipartition* of the magnetic sites. One can show that, for a given crystallographic space group $G$, each bipartition corresponds to a *spin group* $S \subset G \times C_2$.§ More generally, co-rotating tensors have the form $T = X(\boldsymbol{\Lambda}) \otimes B$, where $X(\boldsymbol{\Lambda})$ is a tensor-valued polynomial in $\boldsymbol{\Lambda}$, whose indices are known as the 'spin indices'. For example, $T_{klm\,ij\ldots} = \Lambda_k \Lambda_l \Lambda_m B_{ij\ldots}$ or $T_{lm\,ij\ldots} = \varepsilon_{lmk} \Lambda_k B_{ij\ldots}$ (where $\varepsilon_{lmk}$ is the totally antisymmetric Levi-Civita tensor), are co-rotating tensors.

Tensors of the form $T$ above are co-rotating by construction. Moreover, one can show that, for any fixed direction of $\boldsymbol{\Lambda}$, $T$ is totally symmetric by the magnetic point group (MPG) defined by $\boldsymbol{\Lambda}$ and $S$, and can therefore represent a physical observable. In general, co-rotating tensors only span a proper subspace of the MPG-symmetric tensors of the same rank, parity and index symmetry. If an MPG-symmetric tensor does not admit a co-rotating form, then the corresponding response is allowed only in the presence of SOC.

The most important and widely discussed co-rotating properties in altermagnets are the magnetic multipole expansions of the spin textures in reciprocal space, for which the tensor $B$ is an even-ranked, fully index-symmetric tensor. Other well-known properties described by tensors of the same form are piezomagnetism, the spin-splitter effect and linear magnetic birefringence.

By contrast, weak ferromagnetism, the anomalous Hall effect, gyrotropic magneto-optical Kerr effect and XMCD are *always* non-co-rotating in altermagnets. For example, a possible form of the weak magnetisation (observed in α-$Fe_2O_3$) is $\boldsymbol{M} = \boldsymbol{L} \times \boldsymbol{D}$ or $M_i = \varepsilon_{ijk} L_j D_k$ where $\boldsymbol{D}$ is the Dzyaloshinskii–Moriya vector. Although $L_j D_k$ is co-rotating, the Levi-Civita symbol in $\varepsilon_{ijk} L_j D_k$ contracts the indices of $\boldsymbol{D}$ and $\boldsymbol{L}$, making the weak magnetisation non-separable, and hence non-co-rotating. The same applies to the XMCD vector, $\boldsymbol{T}$, obtained from the anti-symmetric part of the multi-site averaged 1st and 3rd order Faraday tensors, discussed in Equations 1,2 in the main text and the SI. Hence, these non-co-rotating properties require SOC to be activated and are expected to be strongly modulated within a given SG by changing the direction of $\boldsymbol{\Lambda}$ and, consequently, the specific MPG within the same SG family.

**Theoretical modelling:** Density functional theory combined with multiplet ligand-field theory (DFT+MLFT) was adopted for the simulation of XMCD spectra.[48] The DFT calculation was performed using the full-potential local-orbital (FPLO) package with a k-mesh of 30×30×30, with which the Fe-O cluster (cutoff radius ~2.22 Å) was defined by Fe 3d and ligand O 2$p$ orbitals. Such *ab initio* cluster calculation provides useful local properties of correlated system, which were subsequently utilized to construct the many-body atomic Hamiltonian addressing the ligand-metal charge transfer effect:

$$H_{\mathrm{CTM}} = H_U^{dd} + H_U^{pd} + H_{l\cdot s}^{d} + H_{l\cdot s}^{p} + H_o^{d} + H_o^{p} + H_o^{L} + H_{\mathrm{hyb}}^{dL} + H_{\mathrm{ex}}.$$

Here, $H_U^{dd}$ and $H_U^{pd}$ represent the Coulomb interactions between the 3d electrons of Fe, and the Fe 2p and 3d electrons, respectively. $H_{l\cdot s}^{d}$ and $H_{l\cdot s}^{p}$ define the SOC effects of Fe 3d and 2p states, respectively. $H_o^{d}$, $H_o^{p}$ and $H_o^{L}$ denote the on-site Hamiltonians of Fe 3d, 2p and ligand O $2p$ orbitals, respectively. The hybridization interaction between Fe 3d and the ligand orbitals is expressed as $H_{\text{hyb}}^{dL}$, and $H_{\text{ex}}$ defines the Weiss field acting on Fe. In this work, we set the Weiss fields (in-plane) along the Néel vector and perpendicular to it to 0.1 and $10^{-4}$ eV, respectively, which yields spin moments of approximately 4.71 and 0.004 $\mu_B$ along the corresponding directions. Moreover, the angular dependence was obtained by projecting $H_{\text{ex}}$ onto the corresponding magnetic configurations.

§ More precisely, given a crystallographic space group $G$, a *Wyckoff orbit* $X$ is a set of crystal positions generated from a single position $x_0$ by application of all elements $g$ of $G$. A *compatible irrep* $\Gamma$ is a 1-dimensional real irreducible representation of $G$ such that, for any $g \in G$ such that at least one element $x$ of $X$ is left invariant ($g\,x = x$), the character $\chi_\Gamma(g)=1$. One can show that $\Gamma$ induces a *bipartition* of the Wycoff orbit, i.e., one can assign ±1 to each element of $X$, to within a global sign change (this is well defined because of the compatibility condition). $\Gamma$ also induces a homomorphism $g \mapsto (g, \chi_\Gamma(g))$ from $G$ into $G \times C_2$, whose image defines a spin group $S \subset G \times C_2$. The tensor $B$ discussed in the text is obtained by projecting a general tensor of given rank and index symmetry onto the $\Gamma$-sector. Note that, in an antiferromagnet, both $\boldsymbol{\Lambda}$ and $B$ are odd by global exchange of the black/white subsets, so the tensor T is well defined if $X(\boldsymbol{\Lambda})$ is odd-ranked in $\boldsymbol{\Lambda}$.

The XMCD spectroscopy was then calculated based on the equation

$$G_{\pm}(\omega) = \langle\psi|T_{\pm}^{\dagger}\frac{1}{(\omega + \mathrm{i}\Gamma/2 + E_i - H_{CTM})}T_{\pm}|\psi\rangle.$$

In specific, the XMCD signal is given by the difference between the responses brought by right ($T_+$) and left ($T_-$) circularly polarized light. Notably, the contributions of all Fe atoms were fully accounted for in the simulation of the XMCD spectra and in the subsequent comparison with experiments.

Spin-resolved density of states (DOS) calculations were performed using the Vienna Ab initio Simulation Package (VASP) within the projector augmented-wave (PAW) framework.[49,50] The exchange–correlation functional was treated using the generalized gradient approximation in the Perdew–Burke–Ernzerhof (PBE) functional form.[51] A Γ-centered 10x10x10 $k$-point mesh was employed for Brillouin-zone sampling, with an energy convergence criterion of $10^{-6}$eV. The DFT + U method was applied on Fe $d$ orbitals with $U = 5.5$ eV and $J = 0.5$ eV.[11] The total spin-resolved DOS of the unit cell was obtained by summing the atom-projected DOS contributions.

## Experiments

**Materials and sample preparation:** Two types of $\alpha$-$Fe_2O_3$ samples were studied in this work: bulk single crystals for spectroscopy and epitaxial thin films for spectroscopy and microscopy, as specified in the text. Bulk single crystals were purchased from Goodfellow. Single crystals with different crystallographic surface orientations, namely $c$-cut $(0001)$, $a$-cut $(\bar{1}2\bar{1}0)$, and $m$-cut $(10\bar{1}0)$, were used to investigate the orientation of the XMCD vector for various MPGs. The epitaxial thin films were grown on a $(0001)$-oriented $\alpha$-$Al_2O_3$ single crystal substrates by pulsed laser deposition using a KrF excimer laser (248 nm). The thin films (~7-10 nm) were deposited at 600 °C, under an oxygen partial pressure of $2.67\times10^{-3}$ mbar, and laser repetition rate of 3 Hz. Finally, the samples were gradually cooled in a high-oxygen-pressure environment to minimise oxygen vacancies formed during the growth. The thin films were prepared from a Rh-doped $\alpha$-$Fe_2O_3$ ($\alpha$-$Fe_{1.97}Rh_{0.03}O_3$) target, where the mild Rh-doping was used to make the Morin transition accessible. Further details can be found in our previous work.[29,37,42] We deposited a conductive Pt overlayer on all samples to facilitate electron-yield detection in XMCD and microscopy experiments at room temperature using a standard sputter coater. We confirmed the quality of our samples by analysing the X-ray absorption spectra and XMLD. The former confirmed the chemical quality, whereas the latter exhibited the expected quadratic dependence of the Néel vector in the basal plane (see Extended Data Figure 6). Moreover, our XMCD measurements confirmed that thin films and single crystals exhibited similar dichroic signatures (Extended Data Figure 4).

**Structural and magnetic characterisation:** The structural characterisation of samples was carried out using a Rigaku SmartLab X-ray diffractometer. 2θ-ω scans, rocking curves (ω scans) and φ scans were used to determine the exact crystallographic orientations and crystalline quality of the sample, and to confirm the in-plane epitaxial alignment of the thin films relative to the substrate. In all cases, the samples did not have any structural twinning. Furthermore, magnetometry measurements were performed using a Quantum Design MPMS magnetometer. The magnetization as a function of temperature ($M$-$T$) was recorded to determine the Morin transition temperature. An in-plane magnetic field of 50 mT was applied during cooling for state preparation and held during the subsequent warming and cooling cycles. In bulk crystals the transition is at ~ 260 K, whereas in thin films it can be tuned as a function of thickness, strain and doping.[29,37,42]

**X-ray absorption spectroscopy (XAS) and circular dichroism:** The XAS and XMCD experiments were conducted at ALBA (BL29-BOREAS) and the Diamond Light Source (I10 and I06-1). Fe L-edge spectra (695-745eV) were acquired using linearly and circularly polarized X-rays in TEY and FY modes. The experimental geometry allowed continuous variation of the X-ray incidence angle by rotating the sample holder from normal ($\theta = 0°$) to grazing ($= 75°$) incidence relative to the sample surface normal. This control of geometry enabled selective sensitivity to different projections of the XMCD vector as seen in Figure 3 and 4. At I06-1 and BL29-BOREAS, magnetic fields of up to 2 T were applied using superconducting vector magnets, allowing full 3-dimensional control over the direction of the field. As discussed in Figures 2,3,4, these fields were used to control orientation of $\boldsymbol{L}$, enabling access to each of the respective

MPGs. Moreover, at I10, fields were applied parallel or antiparallel to the X-ray propagation direction using either a superconducting high-field magnet or an electromagnet, with maximum field strengths of 14 T and 2 T, respectively. These high fields were used to capture the evolution of the XMCD from the canted moment, see Figure 4.

**X-ray photoemission electron microscopy:** To image the antiferromagnetic domain structure of c-cut α-$Fe_2O_3$ thin films, Fe L-edge X-PEEM measurements were conducted at the I06 beamline, Diamond Light Source. Samples were mounted in an ultrahigh-vacuum chamber with a base pressure of $\sim 6 \times 10^{-10}$ mbar on a customised temperature-controlled cartridge, allowing imaging across the Morin transition. X-rays were incident in a grazing geometry of 16° relative to the sample surface, and the emitted secondary electrons were collected by the electron microscope to form X-PEEM images.

XMLD-PEEM images were acquired by recording, at a fixed linear polarisation (either LH or LV), two X-PEEM images at photon energies ($E_1 = 708.15$ eV and $E_2 = 709.15$ eV), selected from two opposite peaks in the Fe $L_3$-edge XMLD spectrum. XMLD-PEEM contrast was then obtained through the asymmetry: $\Delta = (I_{E_1} - I_{E_2})/(I_{E_1} + I_{E_2})$. This contrast is sensitive to the local orientation of the Néel vector through the angular dependence of the linear dichroic absorption. For $T > T_{\mathrm{M}}$, where the Fe spins lie in the sample plane, the XMLD-PEEM intensity follows $I = I_A + I_B cos^2 \Psi$, where $\Psi$ is the angle between the X-ray polarisation and the local spin axis.[29,30] Domains with different in-plane Néel-vector orientations therefore produce different XMLD contrast. Because XMLD depends quadratically on the magnetic orientation, antiparallel Néel vectors remain indistinguishable, resulting in a 180° ambiguity. For $T < T_{\mathrm{M}}$, the spins reorient out of plane, so the XMLD-PEEM contrast becomes only visible in domain walls separating upward- and downward-oriented domains.

XMCD-PEEM images were acquired at a fixed photon energy of 722.6eV, corresponding to a pronounced peak in the Fe $L_2$-edge XMCD spectrum, using right- and left-circularly polarised X-rays. XMCD-PEEM contrast was calculated as $\delta = (I_{RCP} - I_{LCP})/(I_{RCP} + I_{LCP})$. XMCD is odd under reversal of the magnetic projection along the X-ray propagation direction, so it provides sign-sensitive contrast and can distinguish between opposite Néel-vector orientations, thereby resolving the 180°ambiguity present in XMLD-PEEM. Moreover, we confirmed that the XMCD contrast exhibits the same sign reversal as observed in the spectral signatures, see Extended Data Figure 7. We used extrinsic defects (non-magnetic, and thus non-dichroic) on the sample surface for focusing, which were subsequently masked out by grey circles or ellipses in the final X-PEEM images.

**XMLD Néel vector maps:** To create XMLD vector maps, the XMLD-PEEM imaging was repeated at six azimuthal sample rotation angles over a range of 180°, in steps of 30°. Data reduction, drift and distortion corrections on the resulting set of energy-asymmetry images were performed using the software Igor Pro. For each pixel in the field of view, the angular dependence of the XMLD contrast was fitted to extract the average in-plane spin direction. The reconstructed in-plane Néel-vector orientations were represented using an R-G-B colour scale. In the easy-axis phase, the in-plane Néel-vector can be found localised only within domain walls. To generate Néel vector maps, we performed individual image normalisation and masking of the out-of-plane Néel-vector domains before fitting. Full details of the XMLD-based reconstruction procedure are described in our previous work.

**Complete vector maps:** To generate the full vector maps, we combined the XMLD vector maps with XMCD at the same location. XMCD-PEEM images were acquired for each location and were corrected for data reduction, drift and distortion using the same process detailed above. As discussed in Figures 2,3, the XMCD couples to the 'effective' magnetic dipole vector $\boldsymbol{T}$ through $\boldsymbol{T} \cdot \boldsymbol{k}$, which is dominated by the normal out-of-plane component, $T_z$. Consequently, even in grazing incidence, the dichroic signal is independent of the X-ray polarisation orientation relative to the crystallographic axes, see Figures 5,6. This was confirmed in our angle-dependent XMCD results (see Extended Data Figure 8). The XMCD image provides information about the sign of the local Néel vector which can be integrated with XMLD vector maps to fully reconstruct the local altermagnetic order. The six colour vector maps seen in Figure 5, were produced by first smoothing and filtering each XMCD dataset to reduce random noise. Light and dark domains, corresponding to opposite Néel-vector orientations, were then separated via thresholding and converted into a binary mask. This mask was then combined with the XMLD reconstruction to provide the full 360° sensitivity to the in-plane Néel-vector orientation.

## Notes

In the final stages of the preparation of this manuscript, we became aware of two other studies exploring XMCD in α-$Fe_2O_3$.[52,53] Our work uniquely exploits three-dimensional spin control through vectorial spectroscopy and microscopy to reveal the origin of XMCD from spin-direction driven local symmetry breaking.

## Acknowledgements

H.J., D.M., Z.W. and C.G. acknowledge the support from the Royal Society URF Grant (URF/R1/241120) and the UKRI Horizon Europe Guarantee Funding (EP/X024938/1). P.G.R. acknowledges the support of the Oxford-ShanghaiTech collaboration project. D.M. was supported by the EPSRC DTP Grant. Part of this work was performed at Diamond Light Source beamlines I10-1 (MM40898-1), I06 and I06-1 (MM41514-1), and at ALBA beamline BOREAS (20250340086). We also acknowledge the Diamond Light Source for provison of facilities in the Materials Characterisation Laboratory which was used for magnetic characterisation. B.Z., R.X. and H.Z. acknowledge the funds by the Deutsche Forschungsgemeinschaft (DFG, German Research Foundation) - Project-ID 443703006 - CRC 1487 Iron, upgraded, and project-ID 463184206 - SFB 1548 FLAIR. Calculations were conducted on the Lichtenberg high performance computer of TU Darmstadt. M.B. acknowledges support from the ERC AdG FRESCO (833973). We thank A. Ariando and the Singapore NRF project (NRF-CRP15-2015-01) for support with materials. We acknowledge the support of Hengli Duan (Diamond) and fruitful discussions with Gerrit van der Laan (Diamond), Jan Masell (Karlsruhe). H.J. dedicates this work in the memory of Professor J.M.D. Coey.

## Author Contributions

P.G.R. and H.J. conceived and led the project. D.M., Z.W., J.-C.L., C.G., performed absorption experiments with inputs from P.G.R., Q.H., P.G., M.V., P.B., L.S.I.V. and P.S. under the supervision of H.J. Imaging experiments were performed by D.M., Z.W., J.-C.L., C.G. and F.M. under the supervision of H.J. and P.G.R. H.J. performed the materials design and growth with support from L.I. and M.B. DFT calculations were performed by R.X. and B.Z. under the supervision of H.Z. D.M., J.-C.L., C.G., and H.J. performed the XRD and magnetometry measurements. D.M. performed the data analysis, including dichroic spectra and vector map reconstructions, using code developed by C.G. under the supervision of H.J. Schematics were produced by D.M. and H.J. P.G.R. developed the theoretical framework of co-rotating properties and the Faraday tensor formalism. H.J. and P.G.R. wrote the manuscript. All authors discussed the results and contributed to the manuscript.

## Extended Data Figures

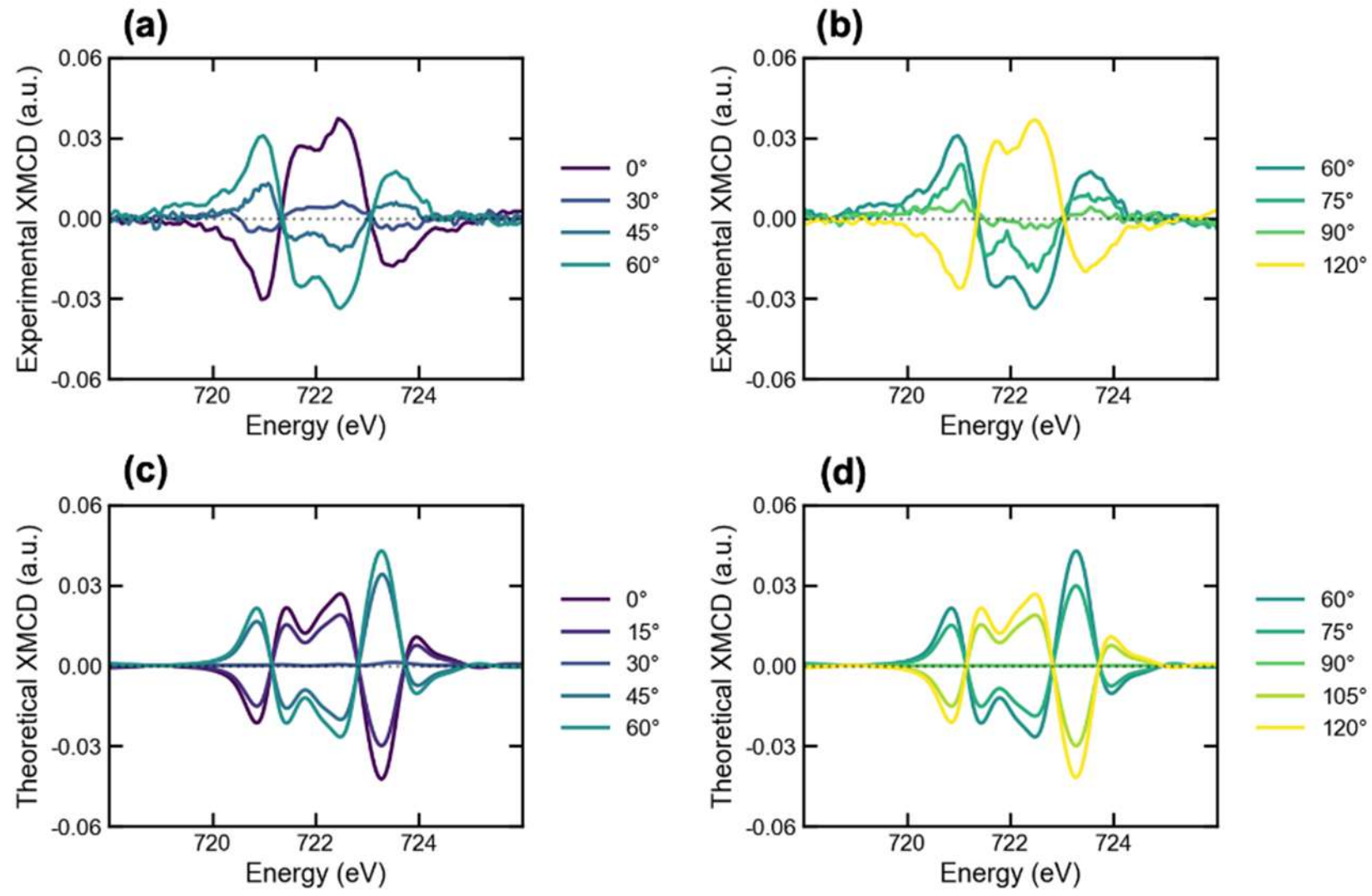

***Extended Data Figure 1 | Fe $L_2$-edge normal-incidence XMCD spectra for in-plane rotation of L. (a-b)*** *Measured (in FY and TEY) and* ***(c-d)*** *calculated XMCD spectra as a function of in-plane rotation angle φ = 0°-120°, where φ defines the angle between* ***L*** *and the a-axis.* ***L*** *was oriented using an in-plane magnetic field of 1 T (FY) and 0.1 T (TEY).*

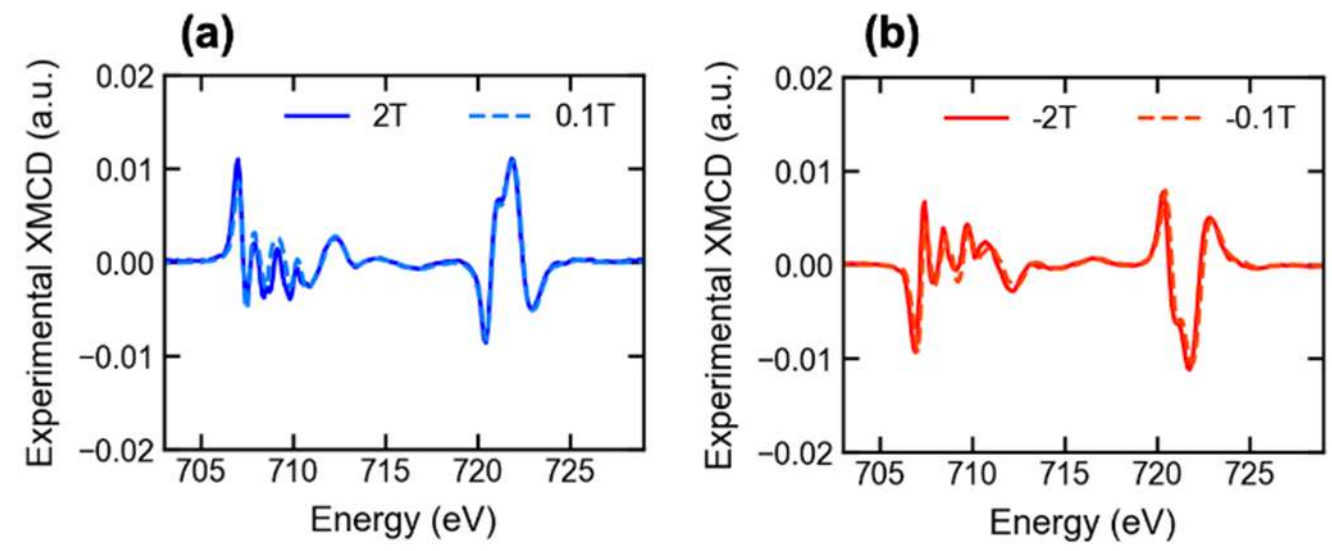

***Extended Figure 2 | Fe $L_2$-edge grazing-incidence XMCD under small applied magnetic fields. (a,b)*** *XMCD spectra, with* ***L*** *along the a-axis, measured under (a) positive and (b) negative applied magnetic fields. No obvious modulation of the lineshape is seen between fields of 2T (solid) and 0.1T (dashed), confirming that in the regime of small applied fields, the 'conventional' XMCD contributed by spin canting can be ignored.*

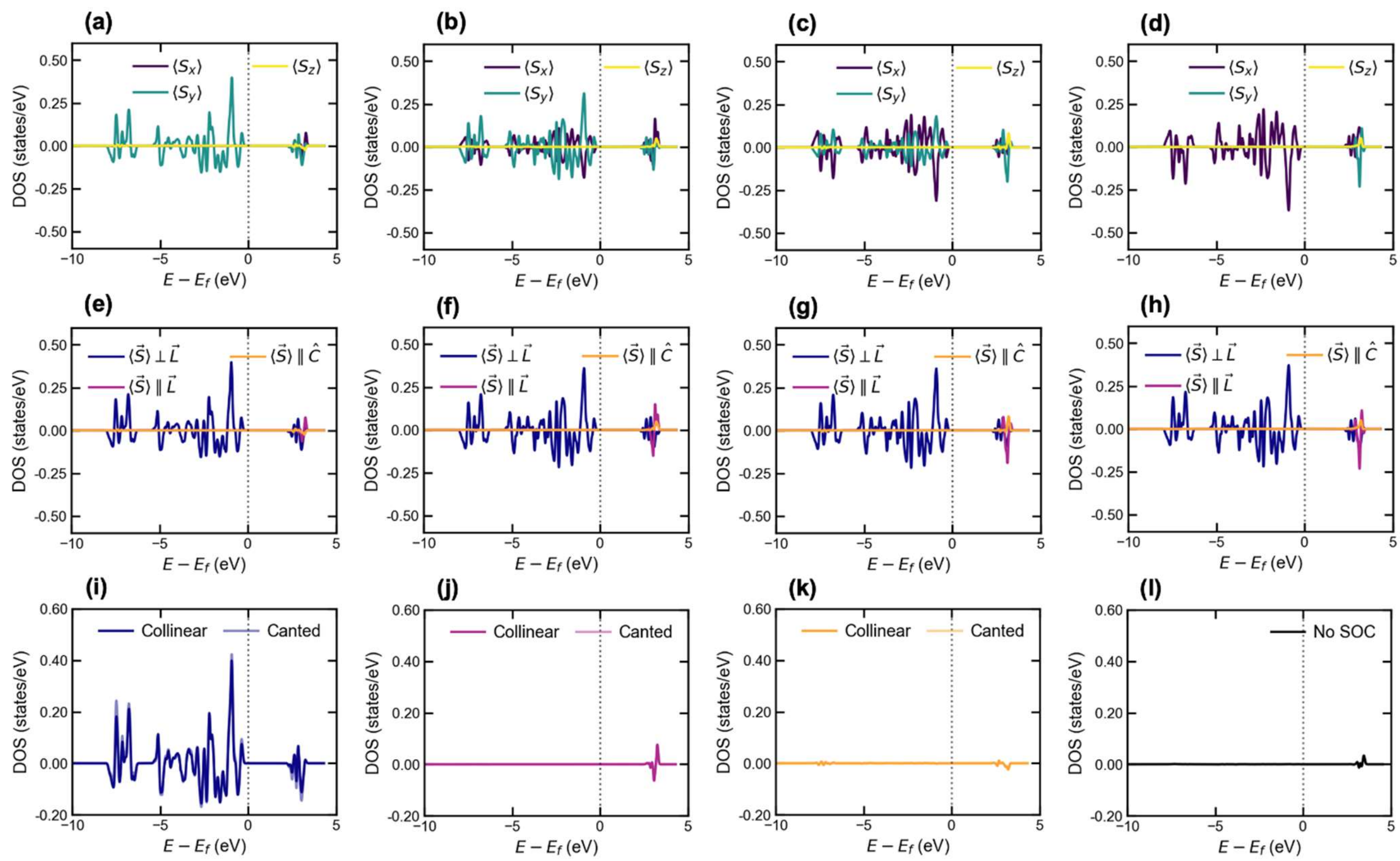

***Extended Figure 3 | Projected spin density of states (DOS) of α-$Fe_2O_3$ for L rotated in the basal plane. (a-h)*** *Calculated projected DOS of α-$Fe_2O_3$ near the Fermi energy for angles φ = (a,e) 0°, (b,f) 30°, (c,g) 60° and (d,h) 90°, where φ is the angle between **L** and the a-axis. The spin DOS are initially shown (a-d) projected onto fixed Cartesian axes (x, y, z) where x is along the* a*-axis and z is along the* c*-axis. As **L** rotates within the plane, this coordinate system remains fixed; as such, under a 90° rotation, $\langle S_x \rangle$ and $\langle S_y \rangle$ are interchanged with a sign inversion. To clarify the dominant orientation of the spin-polarised DOS, in (e-h) we project them on the axes perpendicular and parallel to **L**. The dominant spin polarisation is always orthogonal to **L** in the basal plane and is thus co-aligned with **M**, remaining largely independent of the orientation relative to the crystal structure. The components of the spin polarisation along **L** and* c*-axis are mostly negligible, with minor numerical contributions in the conduction band. **(i-k)** The projected DOS for φ = 0° is shown with and without bulk Dzyaloshinskii-Moriya canting, for the three cases of spin-polarisation (i) orthogonal to **L**, (j) along **L**, and (k) along the* c*-axis. Canting introduces a small increase in the projected DOS, which can be seen in (i) and is negligible in (j) and (k). **(l)** In the absence of SOC, all spin-polarised contributions vanish. Small signals can be attributed to numerical errors in the conduction band. We note that altermagneitc contributions must vanish upon momentum-space integration at fixed energy, because the spin-split bands are symmetric under inversion and contain even nodal planes.*

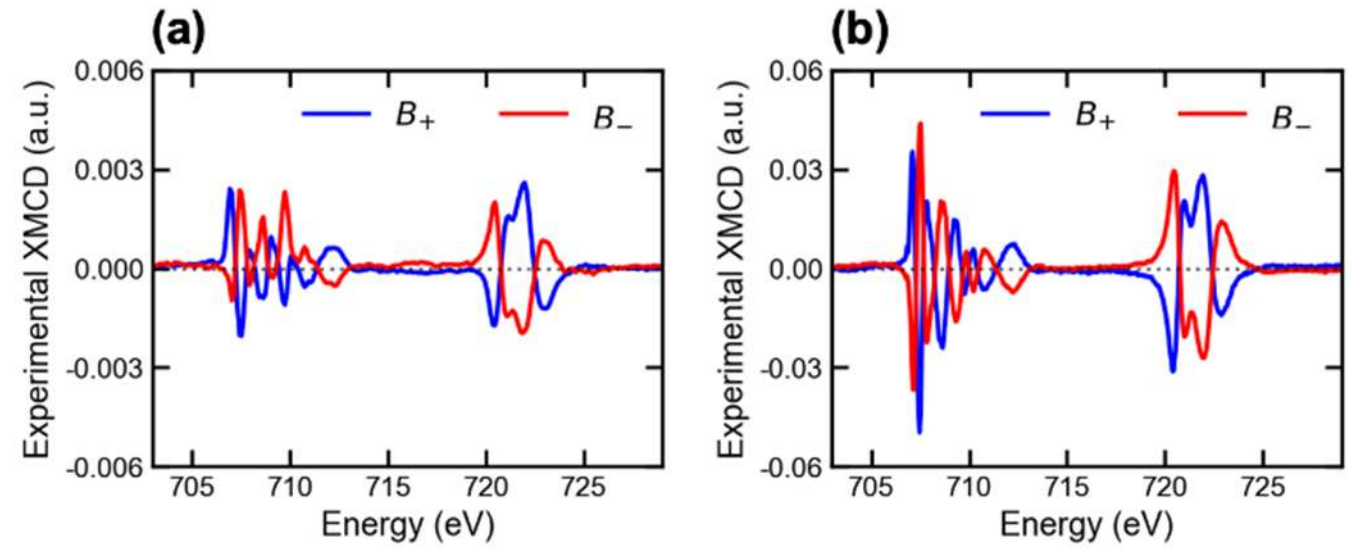

***Extended Data Figure 4 | Fe L-edge normal-incidence XMCD spectra measured in thin film vs single crystal α-$Fe_2O_3$.*** *Measured XMCD (TEY) spectra in a (a)* c*-cut Pt/α-$Fe_2O_3$/α-$Al_2O_3$ thin film and (b)* c*-cut single crystal of α-$Fe_2O_3$. In both samples, **L** was aligned parallel to the* a*-axis. A weaker signal was observed in the thin film due to a smaller fraction of aligned magnetic domains in the small field remnant state.*

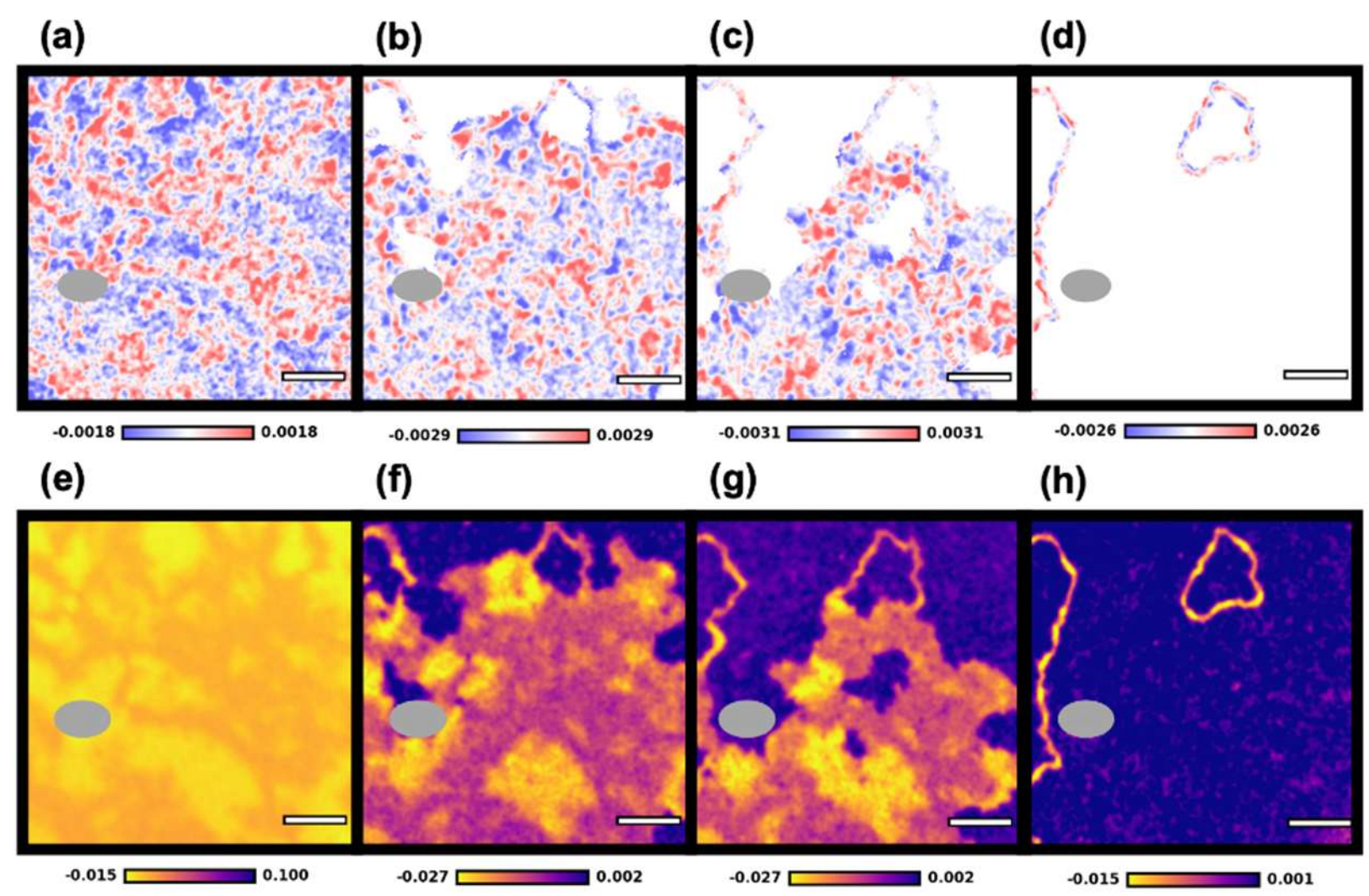

***Extended Data Figure 5 | XMCD-PEEM and XMLD-PEEM imaging of domain reconfiguration through Morin transition. (a-d)*** *XMCD-PEEM and* ***(e-h)*** *XMLD-PEEM images of a region of thin film α-$Fe_2O_3$ shown in the main text Figure 5. Images are shown at (a,e) T = 390 K, (b,f) T = 367 K, (c,g) T = 363 K and (d,h) T = 358 K, across the Morin transition. In (a-d), blue-white-red contrast corresponds to negative, zero and positive $T_z$ domain orientations. In (e-h), yellow and orange colours indicate orthogonal in-plane* ***L*** *orientations, while purple defines out-of-plane* ***L****. Defects are masked in grey. Spatial scale bar is 2µm.*

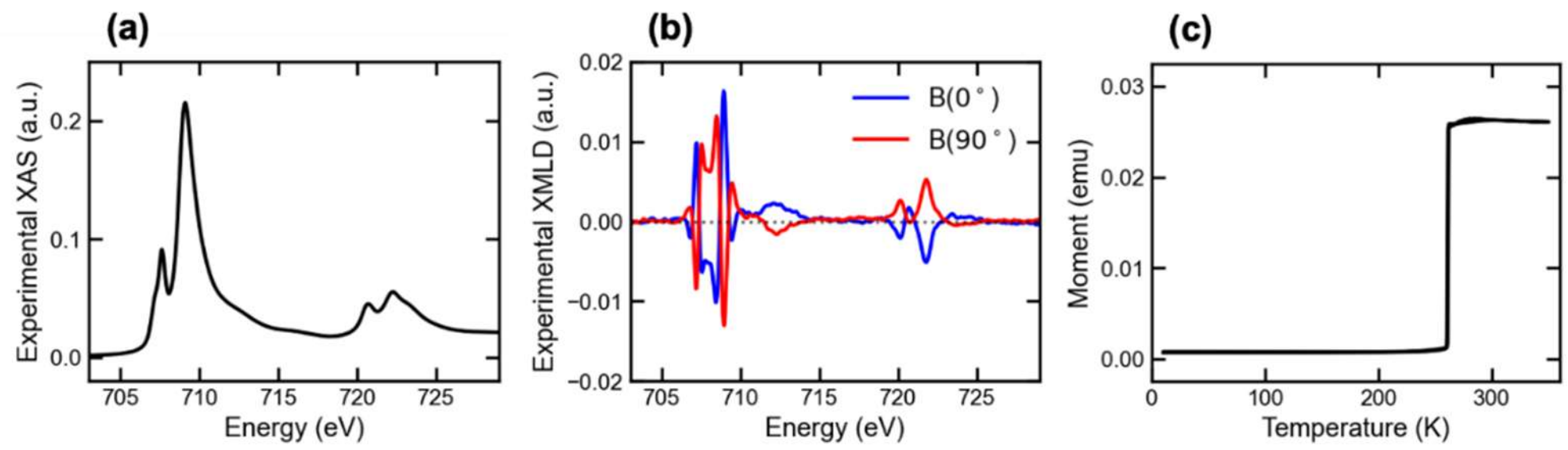

***Extended Data Figure 6 | α-$Fe_2O_3$ magnetic characterisation. (a)*** *XAS of the sample reveals characteristic splitting at the $L_3$ and $L_2$ edges,[29,37] confirming the chemical and phase purity.* ***(b)*** *Measured in-plane XMLD (luminescence yield) for* ***L*** *aligned parallel to the a-axis (blue) and the m-axis (red).* ***(c)*** *Magnetometry data demonstrating the manifestation of the canted moment across the Morin transition in c-cut α-$Fe_2O_3$. In-plane magnetic fields of (a) 0.1 T and (c) 2 T were used to align* ***L****.*

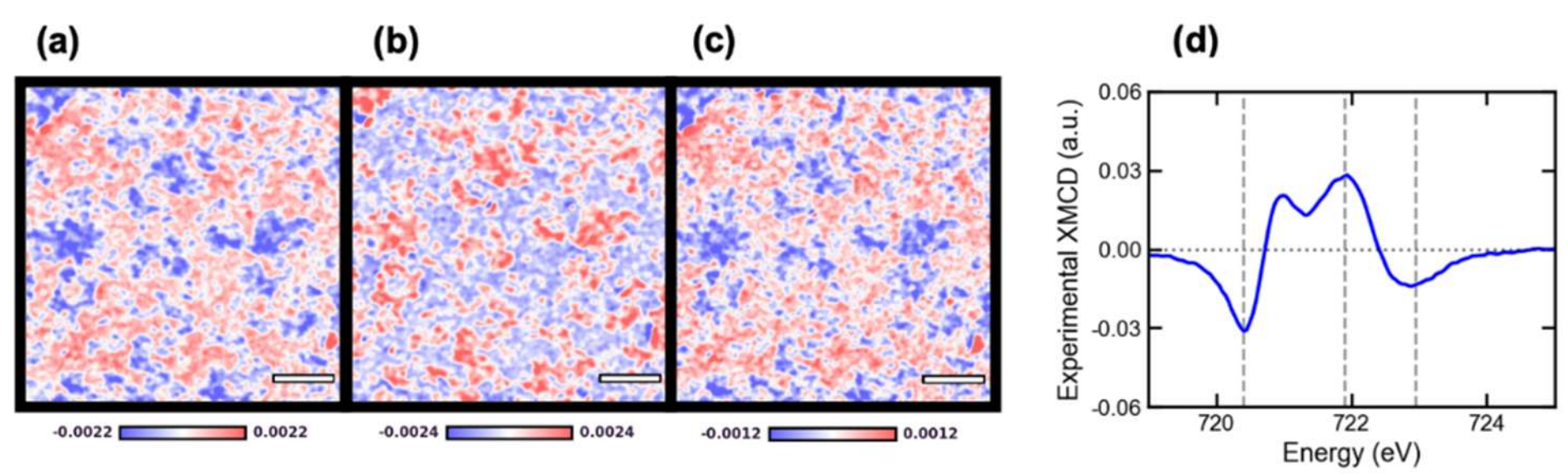

***Extended Data Figure 7 | XMCD contrast reversal across the Fe $L_2$-edge in thin film α-$Fe_2O_3$ above Morin transition.*** *XMCD-PEEM images acquired at three energies across the $L_2$-edge:* ***(a)*** *E = 720.5 eV,* ***(b)*** *E = 721.9 eV and* ***(c)*** *E = 723.0 eV, seen plotted in* ***(d)****. The domain contrast reverses sign across the absorption edge, reflecting the energy-dependent sign change of the XMCD signal. Images were collected at T = 300 K.*

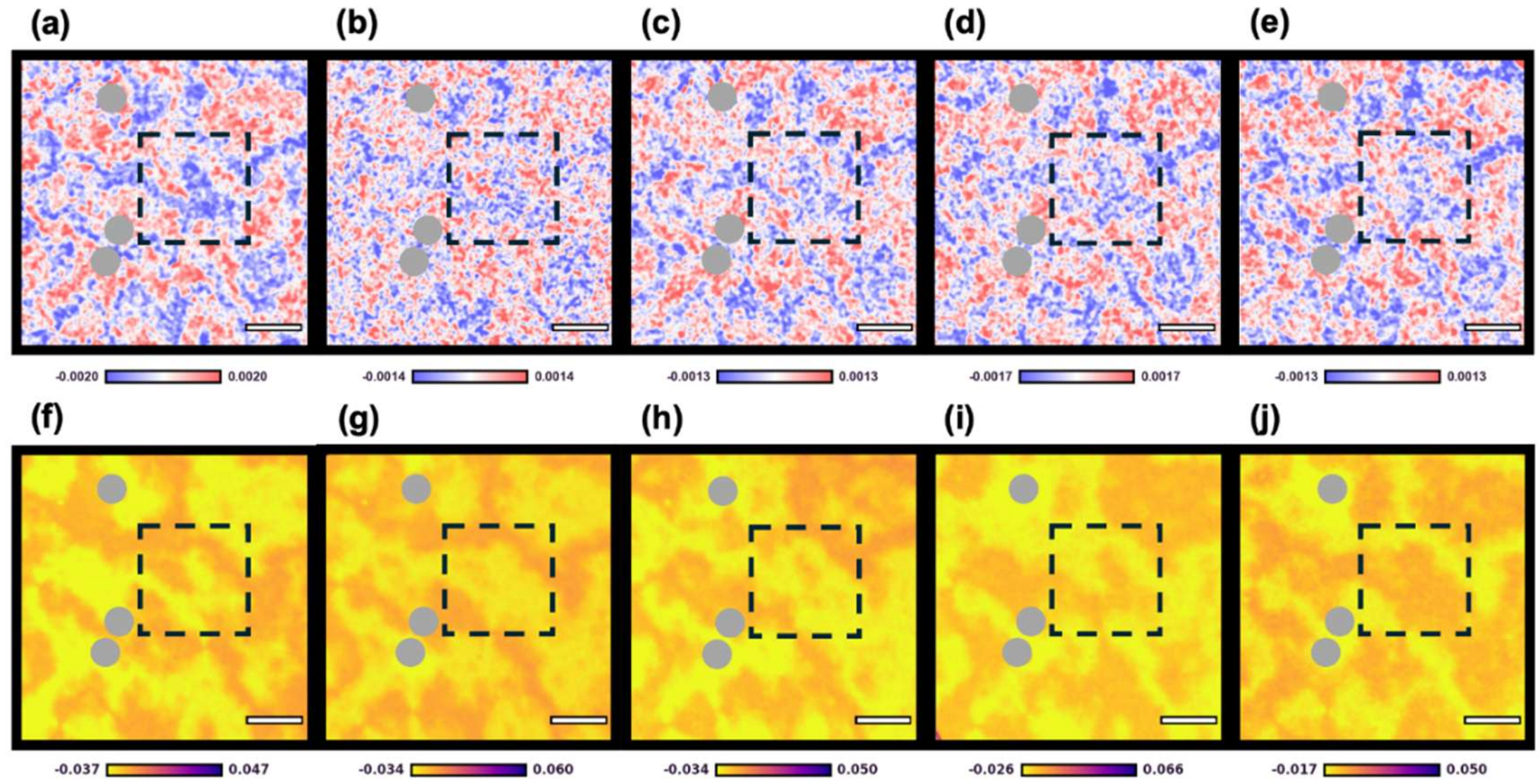

***Extended Data Figure 8 | XMCD- and XMLD-PEEM imaging of magnetic domain structure in thin film α-$Fe_2O_3$. (a-e)*** *XMCD-PEEM and* ***(f-j)*** *XMLD-PEEM images acquired in grazing incidence (16°) for sample rotations (a,f) 0°, (b,g) 30°, (c,h) 60°, (d,i) 90° and (e,j) 120° rotations about the* c*-axis, respectively. The same region is indicated across all images by dashed squares. No significant changes in XMCD domain contrast are observed upon rotation of the incident X-ray direction, consistent with the discussion in the main text. The small variations present are attributed to noise due to the small out-of-plane projection of the X-ray* $\boldsymbol{k}$ *vector. In contrast, the XMLD-PEEM images show a clear evolution of domain contrast with rotation, completely reversing colours between 0° and 90° azimuth orientation, in agreement with previous reports.[29,38] Spatial scale bars are 2 µm.*

## Supplementary Information

### Description of the Faraday tensors

The non-compensated $\mathcal{T}$-odd XMCD vector, $\boldsymbol{T}$, originates from averaging the on-site Faraday tensors associated with the two sublattices, as discussed in Equations 1,2 in the main text. The 1st and 3rd order contributions to the $\boldsymbol{T}$ vector can be extracted from the anti-symmetric part of the site-averaged impermeability tensors $\beta$, through the relations: $T_i^{(1)} = \epsilon_{ijk}\, \beta_{jk}^{(1)\mathrm{avg}}$, $T_i^{(3)} = \epsilon_{ijk}\beta_{jk}^{(3)\mathrm{avg}}$. Specifically, the 1st order Faraday term can be written as the combined effect of both antiferromagnetic sublattices ($a$,$b$), illustrated in Figure 1:

$$\beta_{ij}^{(1)\mathrm{avg}} = \beta_{ij}^{(1),a} + \beta_{ij}^{(1),b} = f_{ijk}^{a} M_k^{a} + f_{ijk}^{b} M_k^{b},$$

where superscripts denote sublattices and not powers. The site symmetry of each Fe-cation is 3. The rank-3 magneto-optical tensor, $f_{ijk}$, is shown below for one of the magnetic sublattices (for simplicity we have dropped the superscript).

$$\begin{pmatrix} \beta_{11} \\ \beta_{12} \\ \beta_{13} \\ \beta_{21} \\ \beta_{22} \\ \beta_{23} \\ \beta_{31} \\ \beta_{32} \\ \beta_{33} \end{pmatrix} = \begin{pmatrix} 0 & 0 & 0 \\ 0 & 0 & f_{123} \\ f_{131} & f_{132} & 0 \\ 0 & 0 & -f_{123} \\ 0 & 0 & 0 \\ -f_{132} & f_{131} & 0 \\ -f_{131} & -f_{132} & 0 \\ f_{132} & -f_{131} & 0 \\ 0 & 0 & 0 \end{pmatrix} \begin{pmatrix} M_1 \\ M_2 \\ M_3 \end{pmatrix}$$

The $f_{ijk}$ tensor for the other sublattice can be obtained by recalling that the two opposite sites are related in hematite by the 2-fold axis. Note that this is an axial tensor that is anti-symmetric under the exchange of the first two indices. It has 3 independent coefficients. The combined contributions from the two sites can be recast in terms of the Néel vector and weak canted moment: $\boldsymbol{M^a} = (\boldsymbol{L} + \boldsymbol{M})/2$, and $\boldsymbol{M^b} = (-\boldsymbol{L} + \boldsymbol{M})/2$. These substitutions can be used to obtain Equation 1 as shown in the main text. The overall response only contains the term confined in the basal plane, orthogonal to the $c$-axis and $\boldsymbol{L}$. Therefore, it has to be collinear with the DMI-allowed canted moment in the system.

Likewise, the 3rd order Faraday term can be written as the combined effect of both antiferromagnetic sublattices ($a$,$b$):

$$\beta_{ij}^{(3)\mathrm{avg}} = \beta_{ij}^{(3),a} + \beta_{ij}^{(3),b} = c_{ijklm}^{a} M_k^{a} M_l^{a} M_m^{a} + c_{ijklm}^{b} M_k^{b} M_l^{b} M_m^{b}$$

where the rank-5 magneto-optical tensor, $c_{ijklm}$, is shown below for one of the magnetic sublattices (for simplicity we have again dropped the superscript). Note that this tensor is anti-symmetric under the exchange of the first two indices, and symmetric under the exchange of the last three indices. The latter property can be used to contract the notation from 5 to 4 indices, $c_{ijklm} \to c_{ij\alpha m}$, by following the usual symmetric contraction, $kl \to \alpha$. Here, $11 \to 1$, $22 \to 2$, $33 \to 3$, $23 = 32 \to 4$, $13 = 31 \to 5$, $12 = 21 \to 6$, and cross terms have a pre-factor 2. Overall, this tensor has 10 independent coefficients.

$$\begin{pmatrix} \beta_{11} \\ \beta_{12} \\ \beta_{13} \\ \beta_{21} \\ \beta_{22} \\ \beta_{23} \\ \beta_{31} \\ \beta_{32} \\ \beta_{33} \end{pmatrix} = \begin{pmatrix}
0 & 0 & 0 & 0 & 0 & 0 & 0 & 0 & 0 & 0 & 0 & 0 & 0 & 0 & 0 & 0 & 0 & 0 \\
-c_{1211} & c_{1211} & 0 & 0 & -c_{1213} & -c_{1212} & -c_{1212} & c_{1212} & 0 & -c_{1213} & 0 & c_{1211} & -c_{1213} & -c_{1213} & -c_{1233} & 0 & 0 & 0 \\
-c_{1311} & -\frac{c_{1311}}{3} & -c_{1331} & -c_{1341} & -c_{1313} & -c_{1312} & -c_{1312} & -3c_{1312} & -c_{1332} & c_{1313} & -c_{1341} & -\frac{c_{1311}}{3} & -c_{1313} & c_{1313} & 0 & -c_{1332} & -c_{1331} & -c_{1341} \\
c_{1211} & -c_{1211} & 0 & 0 & c_{1213} & c_{1212} & c_{1212} & -c_{1212} & 0 & c_{1213} & 0 & -c_{1211} & c_{1213} & c_{1213} & c_{1233} & 0 & 0 & 0 \\
0 & 0 & 0 & 0 & 0 & 0 & 0 & 0 & 0 & 0 & 0 & 0 & 0 & 0 & 0 & 0 & 0 & 0 \\
3c_{1312} & c_{1312} & c_{1332} & c_{1313} & -c_{1341} & -\frac{c_{1311}}{3} & -\frac{c_{1311}}{3} & -c_{1311} & -c_{1331} & c_{1341} & c_{1313} & c_{1312} & -c_{1341} & c_{1341} & 0 & -c_{1331} & c_{1332} & c_{1313} \\
c_{1311} & \frac{c_{1311}}{3} & c_{1331} & c_{1341} & c_{1313} & c_{1312} & c_{1312} & 3c_{1312} & c_{1332} & -c_{1313} & c_{1341} & \frac{c_{1311}}{3} & c_{1313} & -c_{1313} & 0 & c_{1332} & c_{1331} & c_{1341} \\
-3c_{1312} & -c_{1312} & -c_{1332} & -c_{1313} & c_{1341} & \frac{c_{1311}}{3} & \frac{c_{1311}}{3} & c_{1311} & c_{1331} & -c_{1341} & -c_{1313} & -c_{1312} & c_{1341} & -c_{1341} & 0 & c_{1331} & -c_{1332} & -c_{1313} \\
0 & 0 & 0 & 0 & 0 & 0 & 0 & 0 & 0 & 0 & 0 & 0 & 0 & 0 & 0 & 0 & 0 & 0
\end{pmatrix} \begin{pmatrix} M_1M_1M_1 \\ M_2M_2M_1 \\ M_3M_3M_1 \\ 2M_2M_3M_1 \\ 2M_3M_1M_1 \\ 2M_1M_2M_1 \\ M_1M_1M_2 \\ \cdots \\ \cdots \\ 2M_1M_2M_2 \\ M_1M_1M_3 \\ M_2M_2M_3 \\ M_3M_3M_3 \\ 2M_2M_3M_3 \\ 2M_3M_1M_3 \\ 2M_1M_2M_3 \end{pmatrix}$$

The combined contributions at the 3rd order, from the two sites, can be recast in terms of the Néel vector and weak canted moment, $\boldsymbol{M^a} = (\boldsymbol{L} + \boldsymbol{M})/2$, and $\boldsymbol{M^b} = (-\boldsymbol{L} + \boldsymbol{M})/2$. The overall response contains several non-vanishing contributions to $\boldsymbol{T}$. We discuss 2 cases in the following, which can be used to understand the origin of Equation 2 in the main text. We use spherical polar coordinates to indicate the direction cosines of the Néel vector and weak canted moment.

*Case 1: $\boldsymbol{M}$ arbitrary, for $\theta = \pi/2$ (easy-plane spin orientation)*

$\boldsymbol{T}$ component parallel to $L_{x,y}$ : $0$

$\boldsymbol{T}$ component parallel to $\boldsymbol{M}$: $\frac{1}{2}\,(|\boldsymbol{L}|^2 + |\boldsymbol{M}|^2)(c_{1311}\,|\boldsymbol{L}| - 3c_{1312}\,|\boldsymbol{M}|)$

$\boldsymbol{T}$ component parallel to $c$-axis: $\frac{1}{4}\,\left(c_{1211}\,|\boldsymbol{L}|\,(|\boldsymbol{L}|^2 - 3|\boldsymbol{M}|^2) + c_{1212}\,|\boldsymbol{M}|(|\boldsymbol{M}|^2 - 3|\boldsymbol{L}|^2)\right)\cos 3\varphi$

*Case 2: $\boldsymbol{M} = 0$, generic $\theta$ (arbitrary spin orientation)*

$\boldsymbol{T}$ component parallel to $L_{x,y}$ : $\frac{3}{2}\,c_{1341}\,|\boldsymbol{L}|^3 \sin^2\theta \cos\theta \cos 3\varphi$

$\boldsymbol{T}$ component parallel to $\boldsymbol{M}$: $\frac{1}{4}\,|\boldsymbol{L}|^3 \sin\theta\,(2\,c_{1311}\sin^2\theta + 6\,c_{1331}\cos^2\theta + 3\,c_{1341}\sin 2\theta\ \sin 3\varphi)$

$\boldsymbol{T}$ component parallel to $c$-axis: $\frac{1}{4}\,c_{1211}\,|\boldsymbol{L}|^3 \sin^3\theta \cos 3\varphi$